\documentclass[twocolumn]{aastex631}
\usepackage{longtable}

\newcommand{\oiii}{[O\,{\sc iii}]}
\newcommand{\oii}{[O\,{\sc ii}]}
\newcommand{\neiii}{[Ne\,{\sc iii}]}

\submitjournal{ApJL}

\shorttitle{Spectroscopy of $z \simeq 8-10$ galaxies from CEERS}
\shortauthors{Arrabal Haro et al.}

\begin{document}

\title{Spectroscopic confirmation of CEERS NIRCam-selected galaxies at \boldmath{$z \simeq 8 - 10$}}

\correspondingauthor{Pablo Arrabal Haro}
\email{parrabalh@gmail.com}

\author[0000-0002-7959-8783]{Pablo Arrabal Haro}
\affiliation{NSF's National Optical-Infrared Astronomy Research Laboratory, 950 N. Cherry Ave., Tucson, AZ 85719, USA}

\author[0000-0001-5414-5131]{Mark Dickinson}
\affiliation{NSF's National Optical-Infrared Astronomy Research Laboratory, 950 N. Cherry Ave., Tucson, AZ 85719, USA}

\author[0000-0001-8519-1130]{Steven L. Finkelstein}
\affiliation{Department of Astronomy, The University of Texas at Austin, Austin, TX, USA}

\author[0000-0001-7201-5066]{Seiji Fujimoto}
\altaffiliation{Hubble Fellow}
\affiliation{Department of Astronomy, The University of Texas at Austin, Austin, TX, USA}

\author[0000-0003-0531-5450]{Vital Fern\'{a}ndez}
\affiliation{Instituto de Investigaci\'{o}n Multidisciplinar en Ciencia y Tecnolog\'{i}a, Universidad de La Serena, Raul Bitr\'{a}n 1305, La Serena 2204000, Chile}

\author[0000-0001-9187-3605]{Jeyhan S. Kartaltepe}
\affiliation{Laboratory for Multiwavelength Astrophysics, School of Physics and Astronomy, Rochester Institute of Technology, 84 Lomb Memorial Drive, Rochester, NY 14623, USA}

\author[0000-0003-1187-4240]{Intae Jung}
\affiliation{Space Telescope Science Institute, 3700 San Martin Dr., Baltimore, MD 21218, USA}

\author[0000-0002-6348-1900]{Justin W. Cole}
\affiliation{Department of Physics and Astronomy, Texas A\&M University, College Station, TX, 77843-4242 USA}
\affiliation{George P.\ and Cynthia Woods Mitchell Institute for Fundamental Physics and Astronomy, Texas A\&M University, College Station, TX, 77843-4242 USA}

\author[0000-0002-4193-2539]{Denis Burgarella}
\affiliation{Aix Marseille Univ, CNRS, CNES, LAM Marseille, France}

\author[0000-0003-4922-0613]{Katherine Chworowsky}\altaffiliation{NSF Graduate Fellow}
\affiliation{Department of Astronomy, The University of Texas at Austin, Austin, TX, USA}

\author[0000-0001-6251-4988]{Taylor A. Hutchison}
\altaffiliation{NASA Postdoctoral Fellow}
\affiliation{Astrophysics Science Division, NASA Goddard Space Flight Center, 8800 Greenbelt Rd, Greenbelt, MD 20771, USA}

\author[0000-0003-4965-0402]{Alexa M.\ Morales}
\affiliation{Department of Astronomy, The University of Texas at Austin, Austin, TX, USA}

\author[0000-0001-7503-8482]{Casey Papovich}
\affiliation{Department of Physics and Astronomy, Texas A\&M University, College Station, TX, 77843-4242 USA}
\affiliation{George P.\ and Cynthia Woods Mitchell Institute for Fundamental Physics and Astronomy, Texas A\&M University, College Station, TX, 77843-4242 USA}

\author[0000-0002-6386-7299]{Raymond C. Simons}
\affiliation{Department of Physics, 196 Auditorium Road, Unit 3046, University of Connecticut, Storrs, CT 06269, USA}

\author[0000-0001-5758-1000]{Ricardo O. Amor\'{i}n}
\affiliation{Instituto de Investigaci\'{o}n Multidisciplinar en Ciencia y Tecnolog\'{i}a, Universidad de La Serena, Raul Bitr\'{a}n 1305, La Serena 2204000, Chile}
\affiliation{Departamento de Astronom\'{i}a, Universidad de La Serena, Av. Juan Cisternas 1200 Norte, La Serena 1720236, Chile}

\author[0000-0001-8534-7502]{Bren E. Backhaus}
\affiliation{Department of Physics, 196 Auditorium Road, Unit 3046, University of Connecticut, Storrs, CT 06269, USA}

\author[0000-0002-9921-9218]{Micaela B. Bagley}
\affiliation{Department of Astronomy, The University of Texas at Austin, Austin, TX, USA}

\author[0000-0003-0492-4924]{Laura Bisigello}
\affiliation{Dipartimento di Fisica e Astronomia ``G.Galilei'', Universit\'a di Padova, Via Marzolo 8, I-35131 Padova, Italy}
\affiliation{INAF--Osservatorio Astronomico di Padova, Vicolo dell'Osservatorio 5, I-35122, Padova, Italy}

\author[0000-0003-2536-1614]{Antonello Calabr{\`o}} 
\affiliation{INAF - Osservatorio Astronomico di Roma, via di Frascati 33, 00078 Monte Porzio Catone, Italy}

\author[0000-0001-9875-8263]{Marco Castellano}
\affiliation{INAF - Osservatorio Astronomico di Roma, via di Frascati 33, 00078 Monte Porzio Catone, Italy}

\author[0000-0001-7151-009X]{Nikko J. Cleri}
\affiliation{Department of Physics and Astronomy, Texas A\&M University, College Station, TX, 77843-4242 USA}
\affiliation{George P.\ and Cynthia Woods Mitchell Institute for Fundamental Physics and Astronomy, Texas A\&M University, College Station, TX, 77843-4242 USA}

\author[0000-0003-2842-9434]{Romeel Dav\'e}
\affiliation{Institute for Astronomy, University of Edinburgh, Blackford Hill, Edinburgh, EH9 3HJ UK}
\affiliation{Department of Physics and Astronomy, University of the Western Cape, Robert Sobukwe Rd, Bellville, Cape Town 7535, South Africa}

\author[0000-0003-4174-0374]{Avishai Dekel}
\affil{Racah Institute of Physics, The Hebrew University of Jerusalem, Jerusalem 91904, Israel}

\author[0000-0001-7113-2738]{Henry C. Ferguson}
\affiliation{Space Telescope Science Institute, 3700 San Martin Dr., Baltimore, MD 21218, USA}

\author[0000-0003-3820-2823]{Adriano Fontana}
\affiliation{INAF - Osservatorio Astronomico di Roma, via di Frascati 33, 00078 Monte Porzio Catone, Italy}

\author[0000-0003-1530-8713]{Eric Gawiser}
\affiliation{Department of Physics and Astronomy, Rutgers, the State University of New Jersey, Piscataway, NJ 08854, USA}

\author[0000-0002-7831-8751]{Mauro Giavalisco}
\affiliation{University of Massachusetts Amherst, 710 North Pleasant Street, Amherst, MA 01003-9305, USA}

\author[0000-0003-0129-2079]{Santosh Harish}
\affiliation{Laboratory for Multiwavelength Astrophysics, School of Physics and Astronomy, Rochester Institute of Technology, 84 Lomb Memorial Drive, Rochester, NY 14623, USA}

\author[0000-0001-6145-5090]{Nimish P. Hathi}
\affiliation{Space Telescope Science Institute, 3700 San Martin Dr., Baltimore, MD 21218, USA}

\author[0000-0002-3301-3321]{Michaela Hirschmann}
\affiliation{Institute of Physics, Laboratory of Galaxy Evolution, Ecole Polytechnique Fédérale de Lausanne (EPFL), Observatoire de Sauverny, 1290 Versoix, Switzerland}

\author[0000-0002-4884-6756]{Benne W. Holwerda}
\affil{Physics \& Astronomy Department, University of Louisville, 40292 KY, Louisville, USA}

\author[0000-0002-1416-8483]{Marc Huertas-Company}
\affil{Instituto de Astrof\'isica de Canarias, La Laguna, Tenerife, Spain}
\affil{Universidad de la Laguna, La Laguna, Tenerife, Spain}
\affil{Universit\'e Paris-Cit\'e, LERMA - Observatoire de Paris, PSL, Paris, France}

\author[0000-0002-6610-2048]{Anton M. Koekemoer}
\affiliation{Space Telescope Science Institute, 3700 San Martin Dr., Baltimore, MD 21218, USA}

\author[0000-0003-2366-8858]{Rebecca L. Larson}
\altaffiliation{NSF Graduate Fellow}
\affiliation{Department of Astronomy, The University of Texas at Austin, Austin, TX, USA}

\author[0000-0003-1581-7825]{Ray A. Lucas}
\affiliation{Space Telescope Science Institute, 3700 San Martin Dr., Baltimore, MD 21218, USA}

\author[0000-0001-5846-4404]{Bahram Mobasher}
\affiliation{Department of Physics and Astronomy, University of California, 900 University Ave, Riverside, CA 92521, USA}

\author[0000-0003-4528-5639]{Pablo G. P\'erez-Gonz\'alez}
\affiliation{Centro de Astrobiolog\'{\i}a (CAB), CSIC-INTA, Ctra. de Ajalvir km 4, Torrej\'on de Ardoz, E-28850, Madrid, Spain}

\author[0000-0003-3382-5941]{Nor Pirzkal}
\affiliation{ESA/AURA Space Telescope Science Institute}

\author[0000-0002-8018-3219]{Caitlin Rose}
\affil{Laboratory for Multiwavelength Astrophysics, School of Physics and Astronomy, Rochester Institute of Technology, 84 Lomb Memorial Drive, Rochester, NY 14623, USA}

\author[0000-0002-9334-8705]{Paola Santini}
\affiliation{INAF - Osservatorio Astronomico di Roma, via di Frascati 33, 00078 Monte Porzio Catone, Italy}

\author[0000-0002-1410-0470]{Jonathan R. Trump}
\affiliation{Department of Physics, 196 Auditorium Road, Unit 3046, University of Connecticut, Storrs, CT 06269, USA}

\author[0000-0002-6219-5558]{Alexander de la Vega}
\affiliation{Department of Physics and Astronomy, University of California, 900 University Ave, Riverside, CA 92521, USA}

\author[0000-0002-9373-3865]{Xin Wang}
\affiliation{School of Astronomy and Space Science, University of Chinese Academy of Sciences (UCAS), Beijing 100049, China}
\affiliation{National Astronomical Observatories, Chinese Academy of Sciences, Beijing 100101, China}
\affiliation{Institute for Frontiers in Astronomy and Astrophysics, Beijing Normal University,  Beijing 102206, China}

\author[0000-0001-6065-7483]{Benjamin J. Weiner}
\affiliation{MMT/Steward Observatory, University of Arizona, 933 N. Cherry Ave., Tucson, AZ 85721, USA}

\author[0000-0003-3903-6935]{Stephen M.~Wilkins}
\affiliation{Astronomy Centre, University of Sussex, Falmer, Brighton BN1 9QH, UK}
\affiliation{Institute of Space Sciences and Astronomy, University of Malta, Msida MSD 2080, Malta}

\author[0000-0001-8835-7722]{Guang Yang}
\affiliation{Kapteyn Astronomical Institute, University of Groningen, P.O. Box 800, 9700 AV Groningen, The Netherlands}
\affiliation{SRON Netherlands Institute for Space Research, Postbus 800, 9700 AV Groningen, The Netherlands}

\author[0000-0003-3466-035X]{{L. Y. Aaron} {Yung}}
\altaffiliation{NASA Postdoctoral Fellow}
\affiliation{Astrophysics Science Division, NASA Goddard Space Flight Center, 8800 Greenbelt Rd, Greenbelt, MD 20771, USA}

\author[0000-0002-7051-1100]{Jorge A. Zavala}
\affiliation{National Astronomical Observatory of Japan, 2-21-1 Osawa, Mitaka, Tokyo 181-8588, Japan}


\begin{abstract}
We present \textit{JWST}/NIRSpec prism spectroscopy of seven galaxies selected from the Cosmic Evolution Early Release Science Survey (CEERS) NIRCam imaging with photometric redshifts $z_{\mathrm{phot}} > 8$. We measure emission line redshifts $z=7.65$ and 8.64 for two galaxies.
For two other sources without securely detected emission lines we measure $z=9.77_{-0.29}^{+0.37}$ and $10.01_{-0.19}^{+0.14}$ by fitting model spectral templates to the prism data, from which we detect continuum breaks consistent with Lyman~$\alpha$ opacity from a mostly neutral intergalactic medium. 
The presence of strong breaks and the absence of strong emission lines give high confidence that these two galaxies have redshifts $z > 9.6$, but the redshift values derived from the breaks alone have large uncertainties given the low spectral resolution and relatively low signal-to-noise ratio of the CEERS NIRSpec prism data.
The two $z\sim10$ sources observed are relatively luminous ($M_{\mathrm{UV}}<-20$), with blue continua ($-2.3\lesssim\beta\lesssim-1.9$) and low dust attenuation ($A_{V}\simeq0.15^{+0.3}_{-0.1}$); and at least one of them has high stellar mass for a galaxy at that redshift ($\log(M_{\star}/M_{\odot})\simeq9.3^{+0.2}_{-0.3}$).
Considered together with spectroscopic observations of other CEERS NIRCam-selected high-$z$ galaxy candidates in the literature, we find a high rate of redshift confirmation and low rate of confirmed interlopers (8\%).
Ten out of 35 $z > 8$ candidates with CEERS NIRSpec spectroscopy do not have secure redshifts, but the absence of emission lines in their spectra is consistent with redshifts $z > 9.6$.
We find that $z>8$ photometric redshifts are generally in agreement (within their uncertainties) with the spectroscopic values, but also that the photometric redshifts tend to be slightly overestimated ($\langle\Delta z\rangle=0.45\pm0.11$), suggesting that current templates do not fully describe the spectra of very high-$z$ sources. 
Overall, the spectroscopy solidifies photometric redshift evidence for a high space density of bright galaxies at $z>8$ compared to theoretical model predictions, and further disfavors an accelerated decline in the integrated UV luminosity density at $z > 8$.

\end{abstract}


\section{Introduction}
\label{sec:intro}

Understanding galaxy formation and evolution during the first hundreds of Myr in the history of the Universe has been and remains one of the biggest astronomical challenges of the last decades. 
Extensive studies based on deep observations with the \textit{Hubble Space Telescope} (\textit{HST}), the {\it Spitzer Space Telescope}, and the largest ground-based facilities have set constraints on the abundance and physical properties of such early galaxies at redshifts below $z\lesssim9$ \citep[e.g.,][among others]{Ellis2013, McLure2013, Matthee2014, Bouwens2015, Finkelstein2015, Oesch2018, Sobral2018, Stefanon2019, Bowler2020, Finkelstein2022a}.

The advent of \textit{JWST} is quickly revolutionizing the exploration of the early Universe within its first $\sim200$-600 Myr. 
Several works have built up samples of galaxy candidates at $z\sim8-17$ in the first deep \textit{JWST}/NIRCam \citep{Rieke2003, Rieke2005, Beichman2012} imaging available from the Early Release Observations (ERO), Early Release Science (ERS), and treasury programs \citep{Castellano2022, Naidu2022a, Bouwens2023, Finkelstein2022b, Finkelstein2023, Adams2023, Atek2023, Austin2023, Donnan2023, Harikane2023a, Rodighiero2023, Yan2023}. 
Apart from the initial public data, studies from the Guaranteed Time Observations (GTO) projects have also presented other very high-$z$ samples, making use of deeper imaging data \citep{Robertson2023, Perez-Gonzalez2023b, Tacchella2023}.

The abundance and brightness of early galaxies found in these studies seems to be in tension with the predictions from most cosmological models \citep[e.g.,][]{Boylan-Kolchin2023, Ferrara2023, Finkelstein2023, Mason2023, Perez-Gonzalez2023b}. 
However, it is important to keep in mind the caveats associated with broad-band-selected high-$z$ samples \citep[see, e.g.,][]{ArrabalHaro2018}, especially when working with a complex, brand new observing facility such as \textit{JWST} to study a still quite unexplored $z\gtrsim8$ redshift regime.
\cite{Zavala2023}, for instance, 
showed that dust-enshrouded galaxies at lower redshifts ($z\lesssim7$) could be misidentified as very high-$z$ objects.  This can occur when a combination of a strong Balmer break with high dust attenuation and strong nebular emission lines result in spectral energy distributions (SEDs) that resemble the emission dropout signature employed for the selection of Lyman Break Galaxies (LBGs) at high redshift \citep[see, e.g.,][]{Giavalisco2002}.
Indeed, this possibility has been verified by recent  observations using \textit{JWST}/NIRSpec \citep{Jakobsen2022} to study promising $z>10$ candidates from the Cosmic Evolution Early Release Science survey (CEERS; Finkelstein et al., in prep.). One galaxy with a persuasive photometric redshift $z \approx 16$ was instead shown to have $z = 4.9$, with strong line emission affecting flux measurements in several NIRSpec photometric bands and mimicking the SED of a galaxy at much higher redshift \citep{ArrabalHaro2023}.
Moreover, \cite{Bouwens2023} reported discrepancies in high-$z$ samples assembled by independent works from the same public data sets, highlighting the differences in the selection criteria employed and the need for more refined \textit{JWST} instrumental calibrations.

Spectroscopic confirmation of such early galaxies is therefore crucial to validate our current photometric samples, identify possible sources of interlopers in the new $z\gtrsim8$ regime we are starting to explore in detail beyond the \textit{HST} boundaries, and help in refining our high-$z$ selection criteria. 
Only after spectroscopic follow up of a statistically significant fraction of the current $z\gtrsim8$ candidates, will we get a good idea of the reliability of the very high-$z$ photometric samples. That will in turn result in a better characterization of the number density of these early galaxies in the heart of the Epoch of Reionization (EoR).

Among the spectroscopic observations of $z>8$ sources carried out so far, \cite{Williams2023} reported a low-mass highly magnified $z=9.51$ galaxy in Director's Discretionary (DD) NIRSpec observations of the galaxy cluster RX~J2129. On the other hand, \cite{Boyett2023} presented a low-magnification massive $z=9.31$ source with high resolution NIRSpec observations around the Abell 2744 galaxy cluster from the GLASS-JWST ERS program \citep{Treu2022}.
Several $z\gtrsim8$ candidates from CEERS have been also observed with NIRSpec \citep{Fujimoto2023, Heintz2023, Tang2023} with promising confirmation rates.

Confirmation of galaxies with redshifts $z \gtrsim 9.6$  becomes more challenging as strong emission lines like H$\beta$ and \oiii4960,5008 redshift beyond the long wavelength limit of NIRSpec ($\sim5.3$ $\mu$m for the NIRSpec prism). For the relatively high nebular excitation conditions that are typically observed for galaxies at these redshifts \citep[e.g.,][]{Matthee2022, Tang2023, Trump2023}, emission lines at wavelengths bluer than H$\beta$ are usually faint and Lyman~$\alpha$ (Ly$\alpha$) emission is often strongly attenuated by the highly neutral intergalactic medium \citep{Dijkstra2014, Hayes2015}.  However, the low spectral resolution of the NIRSpec prism offers excellent sensitivity for detecting the redshifted ultraviolet continuum emission from galaxies at the EoR, and hence the strong break at Ly$\alpha$  due to the opacity of the intergalactic medium (IGM).

To date, only a few galaxies have been spectroscopically confirmed with NIRSpec at $z > 9.6$. \cite{Roberts-Borsani2022} confirmed a highly-magnified $z=9.76$ galaxy lensed by the Abell 2744 galaxy cluster, and a $z=10.17$ galaxy triply-lensed by the MACSJ0647.7+7015 galaxy cluster was also reported in \citet{Harikane2023b} and \citet{Hsiao2023}. Three galaxies at $z=10.1$, 11.0 and 11.4 have been confirmed by DD observations of the CEERS field \citep{ArrabalHaro2023, Harikane2023b}. Finally, the \textit{JWST} Advanced Deep Extragalactic Survey (JADES) GTO project has presented five others: \cite{Bunker2023} measured an unambiguous redshift of 10.60 for GN-z11, slightly below previous estimates \citep{Oesch2016, Jiang2021}; while \cite{Curtis-Lake2023} presented four other galaxies at $z>10$, including the most distant spectroscopic confirmation at $z=13.2$, identified in deep GTO NIRCam observations \citep{Robertson2023}.

Here we present the spectroscopic confirmation of four $z>8$ galaxies in late CEERS NIRSpec prism observations, including two galaxies at $z\approx9.75-10$. 
This work is structured as follows: \S\ref{sec:data} describes the NIRCam data used to select our targets as well as the NIRSpec observations and data reduction; \S\ref{sec:analysis} shows the redshift measurements and SED fitting of the galaxies; the main results are discussed in \S\ref{sec:results} and summarized in \S\ref{sec:summary}. All magnitudes are in the AB system \citep{Oke1983} and all uncertainty intervals correspond to the 16-84th percentiles of the values. Throughout this work we use a \cite{Planck2020} flat $\Lambda$CDM cosmology with $H_0 = 67.36$ km s$^{-1}$ Mpc$^{-1}$ and $\Omega_{\mathrm{m}} = 0.315$.

\section{Observations and data reduction}
\label{sec:data}

\begin{figure*}
\centering
\includegraphics[width=\textwidth]{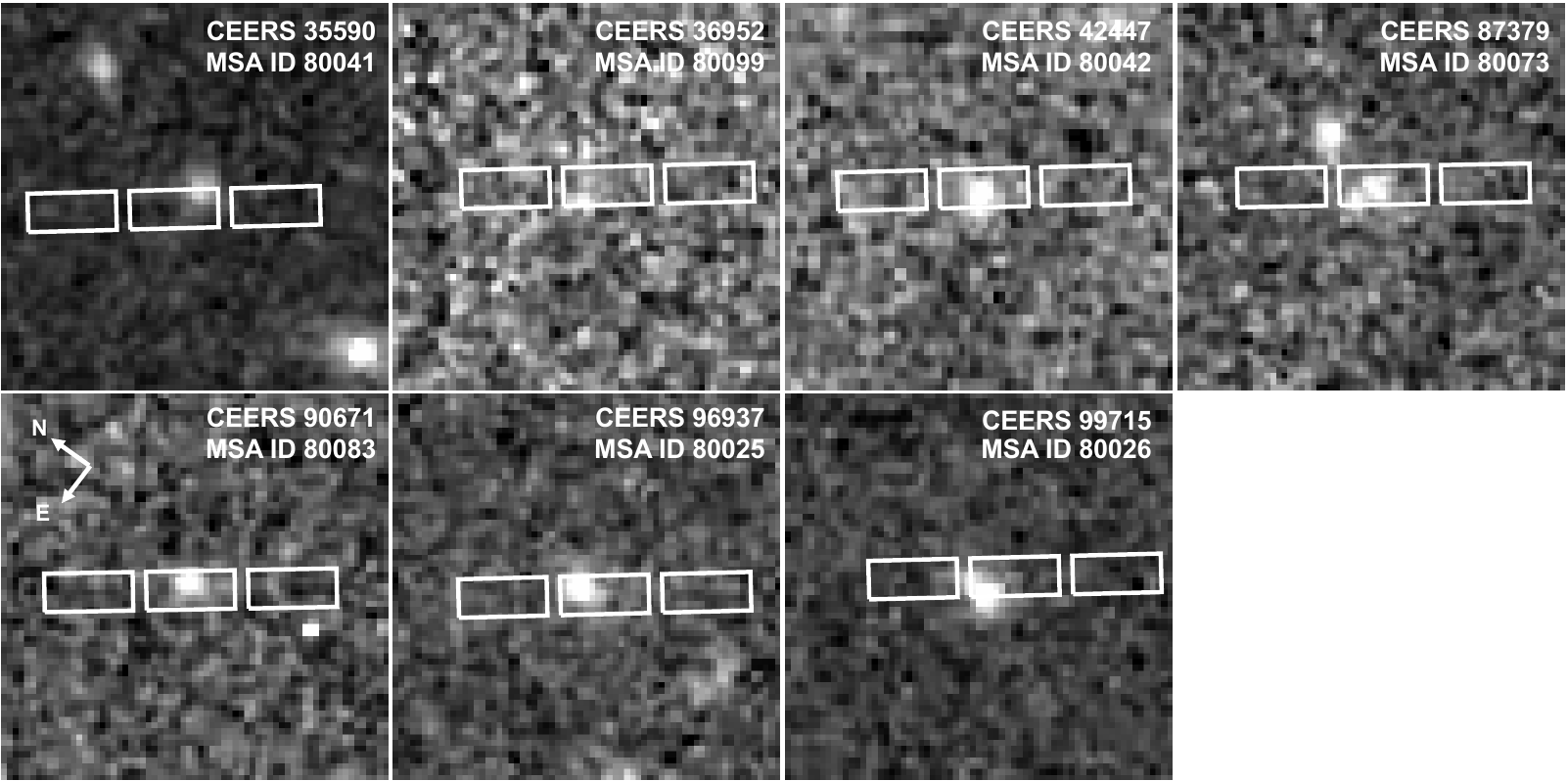}
\caption{F277W 2$''\times2''$ cutouts of the seven NIRCam-selected $z>8$ candidates included in CEERS epoch 3 NIRSpec MSA observations with the NIRSpec MSA shutter positions overlaid.}
\label{fig:full_cutout_mosaic}
\end{figure*}

\subsection{NIRCam data and sample definition}
\label{sec:NIRCam_data}

We selected candidate spectroscopic targets from the CEERS NIRCam imaging data in the CANDELS \citep{Grogin2011, Koekemoer2011} EGS field.  Here we use the full set of 10 CEERS NIRCam pointings, including data obtained in CEERS epoch 2 (December 2022), which were reduced in the same way as the epoch 1 (June 2022) pointings as described in detail in \citet{Bagley2023}.  Photometry was performed on these mosaics in a similar manner as described in \citet{Finkelstein2023}, with a few key differences regarding point spread function (PSF) matching.  To derive accurate total fluxes, all bands with PSFs smaller than that of F277W (i.e., ACS F606W, F814W, and NIRCam F115W, F150W and F200W) had their images convolved to match the F277W PSF.  For images with larger PSFs than F277W (i.e., NIRCam F356W, F410M, F444W, and WFC3 F105W, F125W, F140W and F160W), fluxes were measured in the native images, but a correction was applied.  This correction was derived on a per-source basis as the ratio of the flux in the native F277W image to that of the F277W image convolved to match the PSF in a given image.  In this manner, we derive accurate colors without blurring all images to match F444W (as was done by \citealt{Finkelstein2023}), with the underlying assumption that there are not significant morphological K-corrections from observed $\sim$1--3 or 3--5 $\mu$m.  Finally, accurate total fluxes were estimated via source-injection simulations, here deriving a magnitude-dependent correction factor (from $\sim$ a few percent for bright galaxies, to $\sim$10\% for faint galaxies).

Photometric redshifts were derived using {\sc eazy} \citep{Brammer2008}, employing the default set of 12 Flexible Stellar Population Synthesis (FSPS) templates, along with the additional blue templates designed by \citet{Larson2022} for the $z>8$ universe.  Candidate $z>8$ galaxies were selected in an identical manner as in \citet{Finkelstein2023}, thus we refer the reader there for more details.

\subsection{MIRI imaging data}
\label{sec:MIRI_data}

One of the galaxies in the sample (CEERS\_35590 / MSA ID 80041) falls in the area covered by the CEERS \textit{JWST}/MIRI imaging for field 9, acquired in CEERS epoch 2 (2022 December). MIRI photometry in the F560W and F770W bands extend SED measurement to longer wavelengths than those observed by NIRCam and NIRSpec.   The data were processed following the procedures discussed elsewhere \citep{Papovich22,Yang2023}.  This produced final science, RMS, and weight--map images registered astrometrically to the NIRCam imaging.  The RMS image includes estimates for the Poisson noise, read-out noise, and correlated pixel noise \citep[see][]{Yang2023}.  

We measured photometry in the MIRI imaging following steps in \cite{Papovich22}.  We first matched the image quality of the F560W image to the F770W image and constructed a weighted-sum ``detection image'' of the two. We then measured source photometry using \texttt{Source Extractor} (version 2.19.5, \citealt{Bertin1996}) in ``dual--image'' mode using the detection image and its weight map for object detection and measuring object fluxes and uncertainties on the F560W and F770W images using the parameters in \cite{Papovich22}, and scaling to a total-magnitude using the \texttt{MAG\_AUTO} aperture derived from the F560W + F770W detection image.   

Formally, CEERS\_35590 is detected, albeit weakly, with measured flux values of $f_\nu(\mathrm{F560W}) = 47 \pm 28$~nJy and $f_\nu(\mathrm{F770W}) = 84\pm 29$~nJy.  We include these measurements in the analysis of the galaxy SED below (see \S\ref{sec:SED_fitting}). 

\subsection{NIRSpec observations and MSA target selection}
\label{sec:NIRSpec_data}

The NIRSpec Micro Shutter Array \citep[MSA;][]{Ferruit2022} data here presented come from the rescheduling of two of the original CEERS pointings whose prism observations were affected by an electrical short in CEERS epoch 2 (December 2022). The rescheduled prism observations were executed in February 2023, enabling the selection of new high-$z$ candidates from the CEERS epoch 2 NIRCam images, as described in \S\ref{sec:NIRCam_data}.  High-redshift candidates from the December 2022 CEERS NIRSpec observations are presented in \cite{Fujimoto2023}.

The CEERS epoch 3 NIRSpec pointings (NIRSpec11, NIRSpec12)  followed the same observing configuration as previous CEERS MSA observations, namely, 3 integrations of 14 groups in NRSIRS2 readout mode per visit, for a total exposure time of 3107~s. Three-shutter slitlets were used, enabling a three-point nodding pattern to facilitate background subtraction. The disperser employed was the prism, which covers the wavelength range 0.6 to $5.3\,\mu$m with varying spectral resolution $R \equiv \lambda/\Delta\lambda \approx 30$ at $\lambda = 1.2\,\mu$m to $>300$ at $\lambda > 5\,\mu$m.  The low resolution of the prism at bluer wavelengths aids the detection of faint UV continuum and the Ly$\alpha$ break, while the higher resolution at red wavelengths facilitates detection of redshifted optical rest-frame emission lines.

In addition, the observing configuration in NIRSpec11 was observed twice, introducing a shift of $\sim67$ mas (1/3 of a shutter width) along the dispersion direction in the second visit. This was done to provide a way to test slit losses as a function of source centering within the NIRSpec microshutters. The locations of the two new prism pointings were selected to maximize the yield of NIRCam-selected $z\gtrsim8$ candidates. In order to support the slit loss test, the NIRSpec11 pointing was also constrained to ensure overlap with CEERS NIRCam grism slitless spectroscopic observations.

The final MSA configurations included a total of 7 NIRCam-selected $z>8$ galaxy candidates. Image cutouts of these candidates are shown in Fig.~\ref{fig:full_cutout_mosaic}.

\subsection{NIRSpec data reduction}
\label{sec:NIRSpec_reduction}

The NIRSpec data processing will be explained in detail in Arrabal Haro et al.\ (in prep.). The main steps of the reduction follow those employed in \cite{Fujimoto2023}, \cite{Kocevski2023}, and \cite{Larson2023}, summarized below.

We make use of the STScI Calibration Pipeline\footnote{\url{https://jwst-pipeline.readthedocs.io/en/latest/index.html}} version 1.8.5 and the Calibration Reference Data System (CRDS) mapping 1061, with the pipeline modules separated into three stages.

In stage one (using the \texttt{calwebb\_detector1} pipeline module), we correct for the detector 1/$f$ noise, subtract the dark current and bias, and generate count-rate maps (CRMs) from the uncalibrated images. We modified the parameters of the \texttt{jump} step to gain an improved correction of the ``snowball'' events\footnote{\url{https://jwst-docs.stsci.edu/data-artifacts-and-features/snowballs-and-shower-artifacts}} often seen in the raw data (associated with high-energy cosmic rays).

The resulting CRMs are then processed through stage two using the \texttt{calwebb\_spec2} pipeline module. At this stage, the pipeline creates two-dimensional (2D) cutouts of the slitlets (each made up of three shutters), corrects flat-fielding,  runs background subtraction making use of the 3-nod pattern, executes the photometric and wavelength calibrations, and resamples the 2D spectra to correct distortions of the spectral trace. We adopt the default pipeline slit loss correction implemented in the \texttt{pathloss} step.

In the final stage (using the \texttt{calwebb\_spec3} pipeline module), we combine the images of the three nods, using customized apertures in extracting the one-dimensional (1D) spectra. The custom extraction apertures are determined by visually identifying high signal-to-noise ratio ($S/N$) continuum or emission lines in our targets, features which are easily recognizable in the 2D spectra. In the case where a source is too faint for any robust visual identification, we define a 4-pixel extraction aperture around a central spatial location estimated from the relative position of the target within its shutter, which we derive from the MSA configuration.
Lastly, the 2D and 1D spectra are simultaneously inspected using the Mosviz visualization tool\footnote{\url{https://jdaviz.readthedocs.io/en/latest/mosviz/index.html}} \citep{JDADF2023} in order to mask any possible remaining hot pixels or other artifacts within the images, as well as the detector gap (when present).

The \textit{JWST} pipeline uses an instrumental noise model to calculate flux errors for the extracted spectra.  As described in Appendix~\ref{sec:appendix_noise}, we test these flux errors and rescale them for the effect of interpolation introduced by the pipeline when resampling the data.

\section{Analysis}
\label{sec:analysis}

\subsection{Redshift measurement}
\label{sec:redshift}

\begin{table*}
\begin{center}
\caption{Redshift measurements of CEERS epoch 3 NIRCam-selected candidates at $z > 8$.}
\label{tab:redshift}

\begin{tabular}{lcccccccc}
\hline\hline
MSA ID  &  R.A.  &  Dec.  &  $z_{\rm phot}$ 
 &  $z_{\rm spec}^{\rm [O\textsc{iii}]}$  &  $z_{\rm spec}^{\rm break,MCMC1}$  &  $z_{\rm spec}^{\rm break,MCMC2}$  &  $z_{\rm spec}^{\rm break,EAZY}$  &   $z_{\rm spec}^{\rm EAZY}$ \\
   &   (deg)  &  (deg)  &    &    &    &   & \\
(1)  &  (2)  &  (3)  &  (4)  &  (5)  &  (6)  &  (7)  &  (8)  &  (9) \\
\hline
80041  &  214.732525  &  52.758090  &  $10.15_{-0.42}^{+0.36}$  &  ---  &  $8.71_{-0.05}^{+0.18}$  &  $8.40_{-0.22}^{+0.79}$  &  $9.15_{-0.91}^{+0.83}$  &  $10.01_{-0.19}^{+0.14}$ \\
80026  &  214.811852  &  52.737110  &  $ 9.76_{-0.09}^{+0.60}$  &  ---  &  $9.63_{-0.15}^{+0.20}$  &  $9.74_{-0.33}^{+0.33}$  &  $9.77_{-0.29}^{+0.37}$  &  $10.01_{-0.30}^{+0.18}$ \\
80083  &  214.961276  &  52.842364  &  $ 8.68_{-0.27}^{+0.21}$  &  $8.638_{-0.001}^{+0.001}$  &  ---  &  ---  &  ---  &  $8.64_{-0.02}^{+0.01}$ \\
80025  &  214.806065  &  52.750867  &  $ 8.47_{-0.24}^{+0.15}$  &  $7.651_{-0.001}^{+0.001}$  &  $7.79_{-0.07}^{+0.07}$  &  $7.57_{-0.03}^{+0.02}$  &  $7.82_{-0.18}^{+0.19}$  &  $7.63_{-0.01}^{+0.02}$ \\
\hline \hline
\end{tabular}
\end{center}

\tablecomments{
(1) Source ID in the CEERS MSA observations.
(2) Right ascension (J2000).
(3) Declination (J2000).
(4) Photometric redshift measured as in \cite{Finkelstein2023}.
(5) Spectroscopic redshift derived from \oiii5008.
(6)-(8) Spectroscopic redshift derived from the fit of the Ly$\alpha$ break through the three methods described in \S\ref{sec:redshift}.
(9) Spectroscopic redshift derived from the {\sc eazy}-based methodology applied to the full wavelength coverage of the spectra.
}
\end{table*}

We measure redshifts from the spectra via several methods. A first estimation is performed for the sources with emission lines by Gaussian fitting of the \oiii5008 line. The line fittings and redshift calculation are performed using \textsc{LiMe}\footnote{\url{https://lime-stable.readthedocs.io/en/latest/}} \citep[see][]{Fernandez2023} on the observed frame.

Secondly, we try three separate methods to measure the redshift via the Ly$\alpha$ break:

\begin{enumerate}
    \item MCMC1:  For the first method, we first create a simplified model spectrum which has three free parameters that describe a sharp Ly$\alpha$ break with a power-law spectral slope redward of the break.  These parameters are the redshift, the UV absolute magnitude at 1500 \AA\ ($M_{\mathrm{UV}}$) and the UV spectral slope ($\beta$).  We note that this model intrinsically assumes the break is sharp.  We then include the effect of Ly$\alpha$ damping wing absorption by adding two additional parameters: the neutral hydrogen fraction ($x_{\mathrm{HI}}$) and an ionized bubble radius ($R_{\mathrm{bubble}}$).  This latter component was introduced by \cite{Curtis-Lake2023} to account for the Ly$\alpha$ damping wing and therefore derive more accurate continuum-break-based redshifts in the EoR.

    We derive posterior constraints on these five parameters using an IDL implementation of the \textsc{emcee} \citep{foreman-mackey13} Python code (see \citealt{Finkelstein2019} for details).  This procedure maximizes the likelihood that the model described by these five free parameters matches the observed prism spectrum for a given source.  For each step in the Markov Chain Monte Carlo (MCMC), the intrinsic spectrum is first generated via the draw of $z$, $M_{\mathrm{UV}}$ and $\beta$.  Then the Ly$\alpha$ damping wing absorption is applied following Equation 30 from \cite{Dijkstra2014}, where the damping wing optical depth is primarily dependent on photon frequency, characterized by the velocity offset. This velocity offset is computed at each wavelength as being proportional to the rest-frame difference between that wavelength and Ly$\alpha$. The model spectrum is then smoothed by two pixels to match the approximate resolution of the prism data.  A likelihood is then calculated assuming the uncertainties are Gaussian, and restricting the spectrum to wavelengths below 2500 \AA\ rest-frame for a given redshift (and omitting regions where emission lines are expected).  Results are derived from the posterior distribution on these five parameters from a chain consisting of 10$^5$ steps following a 10$^6$-step burn-in process.

    \item MCMC2:  Similarly, the Ly$\alpha$ break redshifts are also calculated using another version of the redshift estimator that is based on the same methodology described for method 1) but uses a separate MCMC sampler package from \cite{Jung2017}.
    We derive posterior distributions of the five parameters similarly to method 1). 
    We have flat priors for all parameters, and the log-likelihood is $-\chi^2/2$ between a modeled transmitted spectrum and an observed spectrum. An additional 5\% of systemic errors are applied in the $\chi^2$ estimation. We employ the Metropolis-Hastings algorithm \citep{Metropolis1953, Hastings1970} in MCMC sampling. The MCMC sampler checks for sampling convergence in a 10$^{4}$-step burn-in process, and the posterior distributions of the parameters are constructed from 10$^4$ chain steps. The fitting results take the median values of the posterior with 1$\sigma$ uncertainties from central 68\% confidence ranges in the posterior.

    \item ``Modifed \textsc{eazy}'':  We perform a third estimation of the Ly$\alpha$ break redshift using the SED fitting code {\sc eazy}. In this approach, we create a top-hat filter response at each wavelength of the prism spectrum and treat the spectrum as a set of pseudo-narrow-band photometry. We adopt a 2-pixel wavelength range as the filter response width for the top-hat filter response. We then use the convolved 2-pixel photometry and filter files to run {\sc eazy} on the region of the spectrum around the Ly$\alpha$ break ($\lambda_{\mathrm{obs}} < 2.5 \mu$m, hereafter ``\textsc{eazy}-break''). To recover the color space of blue, young galaxies, we utilize the 6 templates from \cite{Larson2022} in addition to the default 12 FSPS templates. An additional redshift estimation (``\textsc{eazy}--full'') is carried out applying this same methodology over the complete spectrum, instead of limited to the wavelengths around the Ly$\alpha$ break.

    It is important to note that with this methodology, no assumptions about the nature of the dropout are made; low-$z$ solutions are considered equally, whereas, for the two methods described above, the assumption of a Ly$\alpha$ break limits the estimations to high-$z$ solutions.
\end{enumerate}

\begin{figure}
\includegraphics[width=\columnwidth]{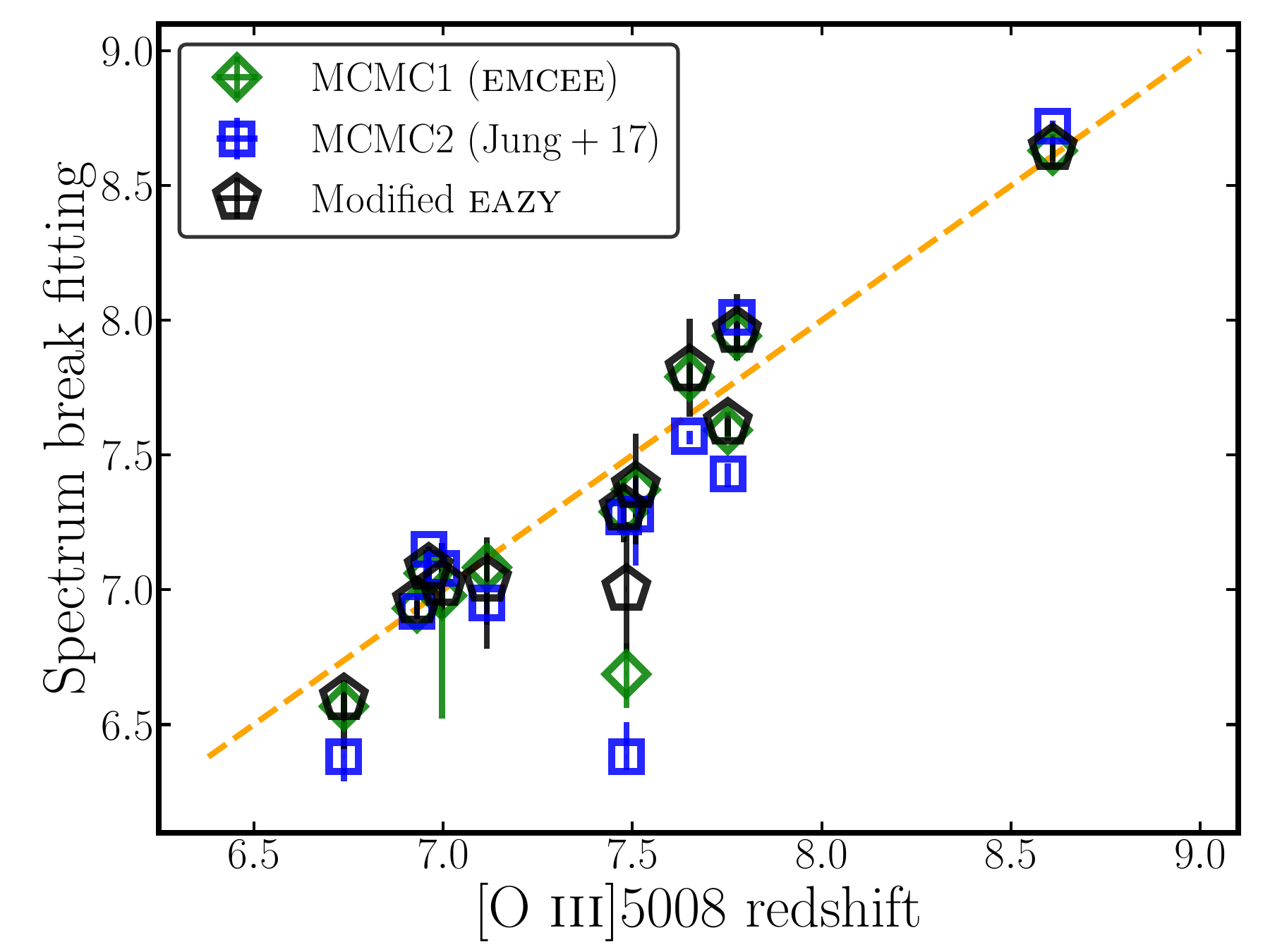}
\caption{Ly$\alpha$ break redshift estimations for CEERS sources at $z > 6.6$ with detections of both the Ly$\alpha$ break and emission lines in NIRSpec prism spectra. 
Several different break redshift measurements are compared to the redshifts derived from the \oiii5008 line, here taken as the fiducial redshifts of the objects.
Three methods are used to fit the Ly$\alpha$ break, as described in the text:  MCMC1 (\textsc{emcee}), MCMC2 \citep{Jung2017}, and modified {\sc eazy}. These are represented by the blue squares, green diamonds and black pentagons, respectively. The one-to-one relation is shown as a dashed orange line.}
\label{fig:zbreak_comparison}
\end{figure}

In order to calibrate the accuracy of the Ly$\alpha$ break redshifts, we apply our three Ly$\alpha$ break fitting algorithms described above to a subsample of 12 galaxies at $6.6 < z < 8.7$ with good detections of both the Ly$\alpha$ break and emission lines in CEERS NIRSpec prism spectroscopy. We compare the resulting Ly$\alpha$ break redshifts with the \oiii5008 redshifts in Fig.~\ref{fig:zbreak_comparison}. The performance of the three methodologies is similar, reporting good redshift estimations ($|\Delta z| / (1+z_{\mathrm{[OIII]}}) < 0.05$) for all sources except one (CEERS\_85358, MSA ID 80372; $z_{\mathrm{[OIII]}} = 7.48$) with faint Ly$\alpha$ emission, whose redshift is more accurately determined by the {\sc eazy}-based approach. The global RMS scatter of $\Delta z$ is 0.18, 0.26 and 0.38 for the {\sc eazy}-based (``modified \textsc{eazy}''), the {\sc emcee} MCMC-based (MCMC1) and the \cite{Jung2017} MCMC-based method (MCMC2), respectively. When omitting the single source with the largest deviation,  RMS($\Delta z$) goes down to 0.12, 0.12 and 0.21 for the same three methods, with an average $\langle\Delta z \equiv z_{\mathrm{break}} - z_{\mathrm{[OIII]}}\rangle$ of -0.05, -0.09 and -0.16, respectively. 
Based on this test, we adopt the redshift value from the {\sc eazy}-based Ly$\alpha$ break fitting when the estimations from the different Ly$\alpha$ break methodologies are consistent.
For reference, we measure RMS($\Delta z$) = 0.76 when comparing the  photometric redshifts against the \oiii-derived value for the galaxies employed in this test.

\begin{figure*}
\centering
\includegraphics[width=0.9\textwidth]{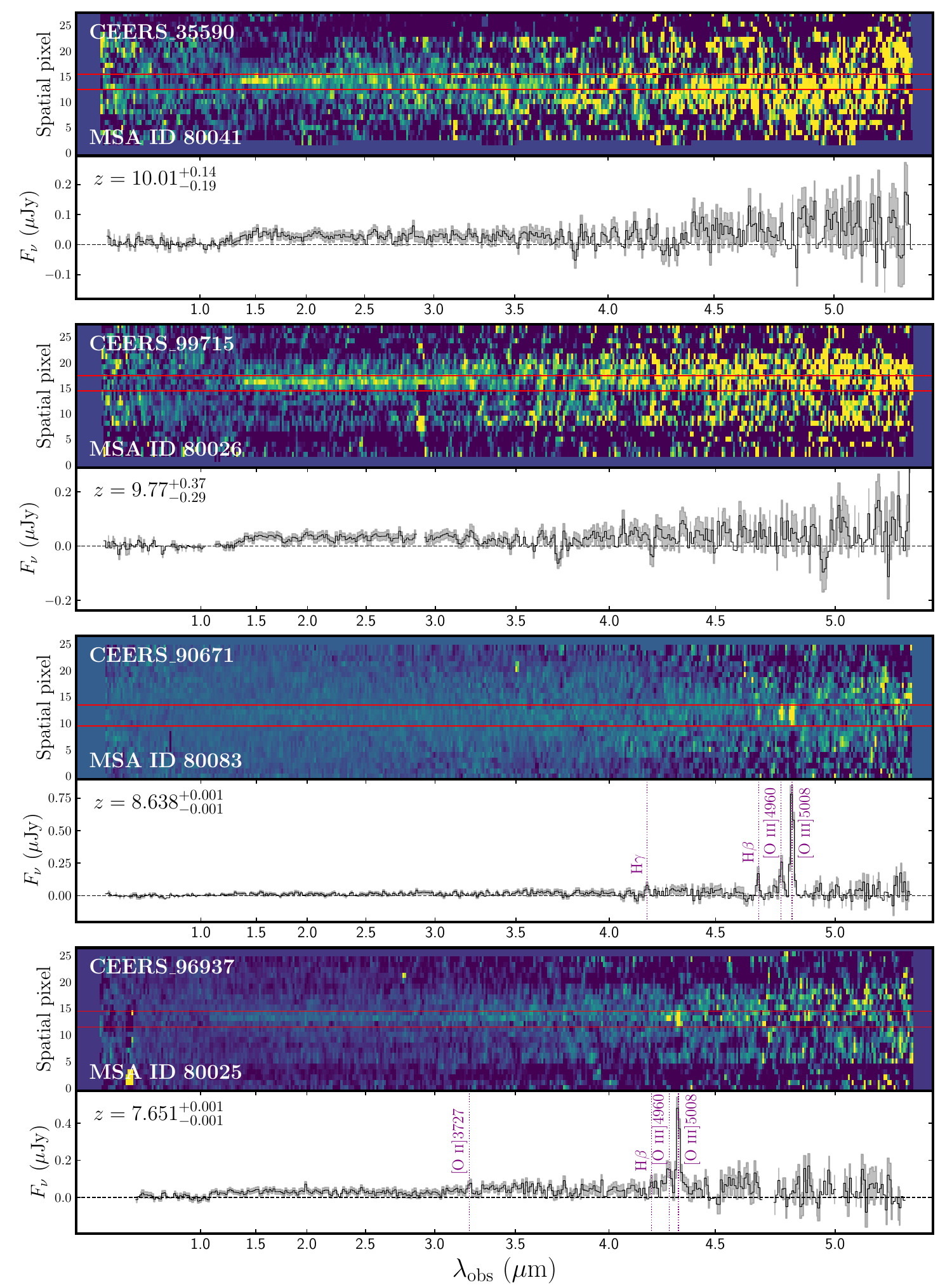}
\caption{NIRSpec prism 2D (\textit{upper panels}) and 1D (\textit{lower panels}) spectra of the confirmed $z\gtrsim8$ galaxies. Emission lines are marked with dotted purple lines when present. The red horizontal lines indicate the limits employed for the 1D extraction.}
\label{fig:spectra_mosaic}
\end{figure*}

Four out of the 7 NIRCam-selected $z>8$ candidates have their continuum and/or emission lines detected with high enough $S/N$ to measure redshifts (see Table~\ref{tab:redshift}). The 2D and 1D spectra of these four sources are presented in Fig.~\ref{fig:spectra_mosaic}. Two galaxies have emission lines from the \oiii\ doublet and (clear or tentative) H$\beta$ at $7.65 \leq z \leq 8.64$. The Ly$\alpha$ continuum break is also detected for one of these galaxies. 
Two other galaxies show continuum breaks near 1.3~$\mu$m without clearly identifiable emission lines.  We discuss these two galaxies in greater detail in the following subsection.

\subsubsection{CEERS\_99715 and CEERS\_35590}

The spectra for galaxies CEERS\_99715 (MSA ID 80026) and CEERS\_35590 (MSA ID 80041) exhibit breaks at $\lambda \approx 1.3\ \mu$m, with little to no flux detected at shorter wavelengths. Integrating the spectra over wavelength intervals 1.35--1.75 $\mu$m and 0.85--1.25 $\mu$m, above and below the breaks, yields  $S/N = 11.8$ and 0.0 for CEERS\_99715 and 12.1 and 1.0 for CEERS\_35590, respectively.  By eye, the break for CEERS\_99715 appears to be ``sharper'', while that for CEERS\_35590 seems more gradual, but given the low $S/N$ per pixel ($<3$ at $\lambda < 1.4\,\mu$m) this apparent difference may not be significant.

For CEERS\_99715, the 2D and 1D spectra show what appears to be an emission line that peaks at the reddest pixel in the spectral range (5.294 $\mu$m), as if the line were truncated by the bandpass cutoff and/or the pipeline processing. In the rectified 2D spectrum the putative line is centered in the cross-dispersion direction within 0.5 pixel of the blue continuum, and visual inspection of the three separate nods shows that it appears to be detected in each. Although the line seems highly significant by eye, the FLUX\_ERROR in the last pixel of the 1D extraction is several orders of magnitude larger than that in the adjacent (bluer) pixel. The elevated error value may be an artifact of the pipeline processing, but it suggests caution interpreting features at the very extremes of the spectral range.

\begin{figure}
\includegraphics[width=\columnwidth]{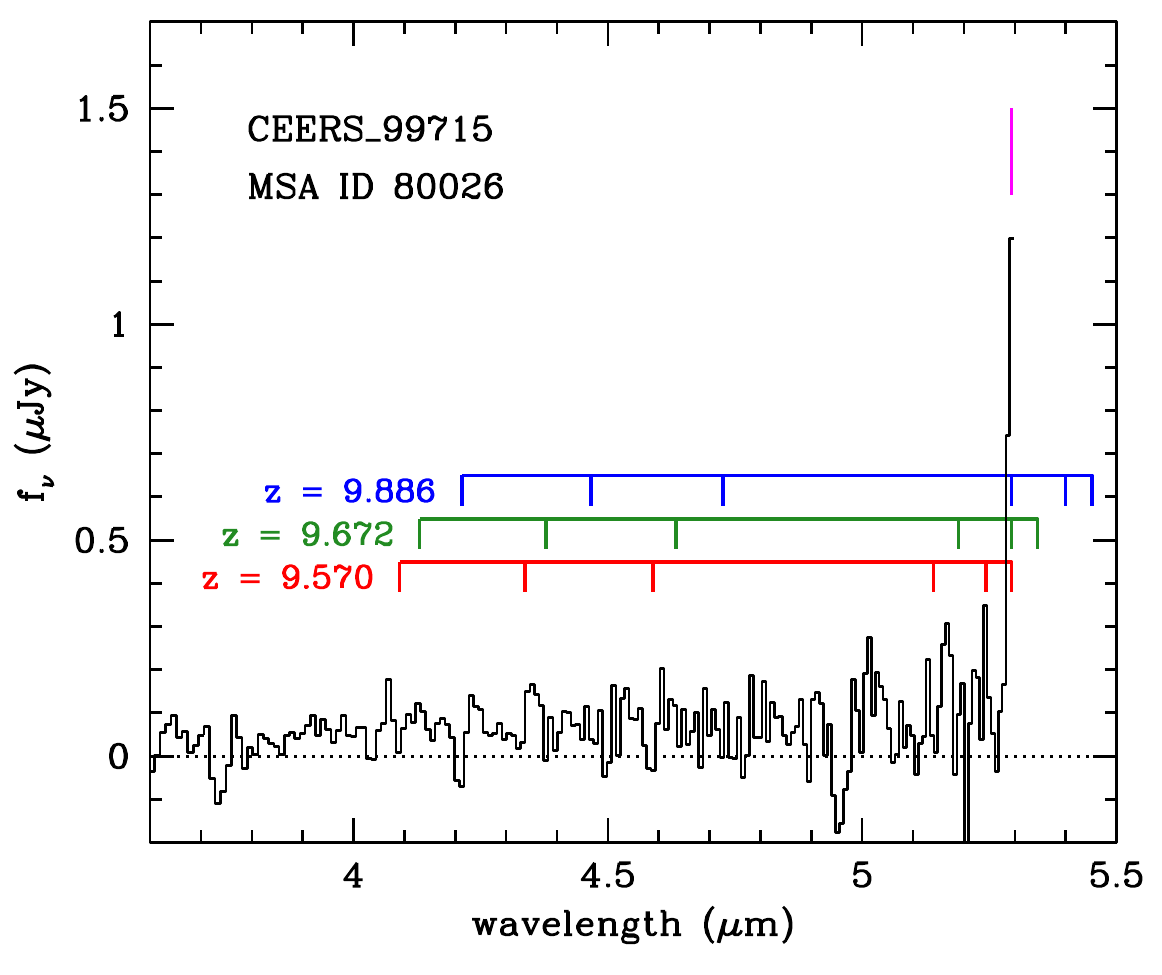}
\caption{The red end of the extracted 1D spectrum for CEERS\_99715. The last two pixels show elevated signal that could be interpreted as a truncated emission line, marked by the magenta line at $5.294\,\mu$m.  The blue, green and red bars indicate the expected locations of emission lines (from left to right: \neiii3870, H$\delta$, H$\gamma$, H$\beta$, \oiii4960 and \oiii5008) at the labeled redshifts under the assumption that the possible line is \oiii5008, \oiii4960 or H$\beta$, respectively.}
\label{fig:tentative_line_options}
\end{figure}

If the apparent emission line in CEERS\_99715 were \oiii5008, \oiii4960, or H$\beta$, the corresponding redshifts would be $z = 9.570$, 9.672 or 9.886, respectively. Given the low spectral resolution of the NIRSpec prism at blue wavelengths, any of these redshifts would be roughly consistent (within 2 pixels) with a Ly$\alpha$ break at $\sim 1.3\ \mu$m. However, in each case we would expect to detect other emission lines in the spectrum (see Fig.~\ref{fig:tentative_line_options}).  If the putative line were \oiii5008 we would expect \oiii4960 at one third the strength \citep{Storey2000}, and perhaps also H$\beta$ and other Balmer lines.  For \oiii4960 we may also expect H$\beta$, and if the line were H$\beta$ we should detect H$\gamma$ unless there is significant nebular reddening, which seems unlikely (unreddened flux ratio H$\gamma$/H$\beta = 0.46$ to 0.47 for Case B recombination at $T = 10,000$~K; \citealt{Osterbrock1989}). No significant emission lines are found at the predicted wavelengths for these lines or others (e.g., \neiii, \oii). 
For other plausible identifications for the apparent 5.294 $\mu$m line (e.g., H$\alpha$) we may also expect other lines that are not seen (like H$\beta$ and \oiii), and the observed Ly$\alpha$ break would start to be inconsistent with the implied $z$ in that scenario.  If the 1.3 $\mu$m break were a Balmer break at $z \approx 2.5$ then the apparent line would have a rest-frame wavelength $\approx 1.5\ \mu$m, where no strong features are expected. Moreover, the possibilities of this line being Pa$\alpha$ or Pa$\beta$ at $z\approx1.8$ or $z\approx3.1$, respectively, would imply a location of the Balmer break that is inconsistent with the measured drop out in the spectrum. We conclude that this apparently strong, truncated emission line at the red limit of the spectrum seems inconsistent with other evidence and may be spurious.

For object CEERS\_35590, a less significant emission feature (peak $S/N = 3.5$ for a 4-pixel extraction; smaller for a narrower extraction) is found at 5.277 $\mu$m, a few pixels short of the red limit of the spectrum.  Following the same reasoning as for CEERS\_99715, no other emission lines are significantly detected at wavelengths predicted under various assumed line identifications, but this is less constraining given the low $S/N$ of the 5.277 $\mu$m feature and its larger deviation from the spatial center of the continuum trace in the 2D spectrum.

\begin{figure*}
\includegraphics[width=\columnwidth]{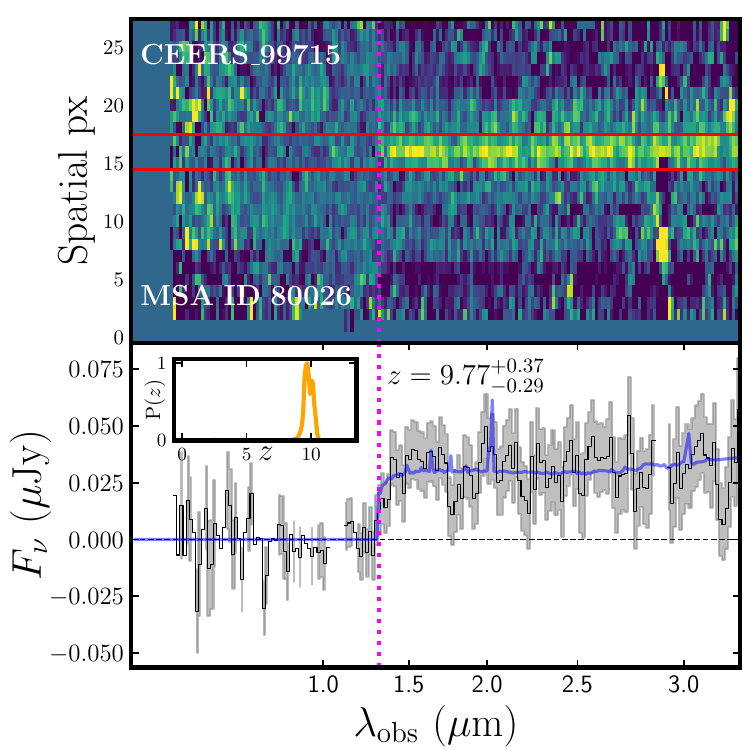}
\includegraphics[width=\columnwidth]{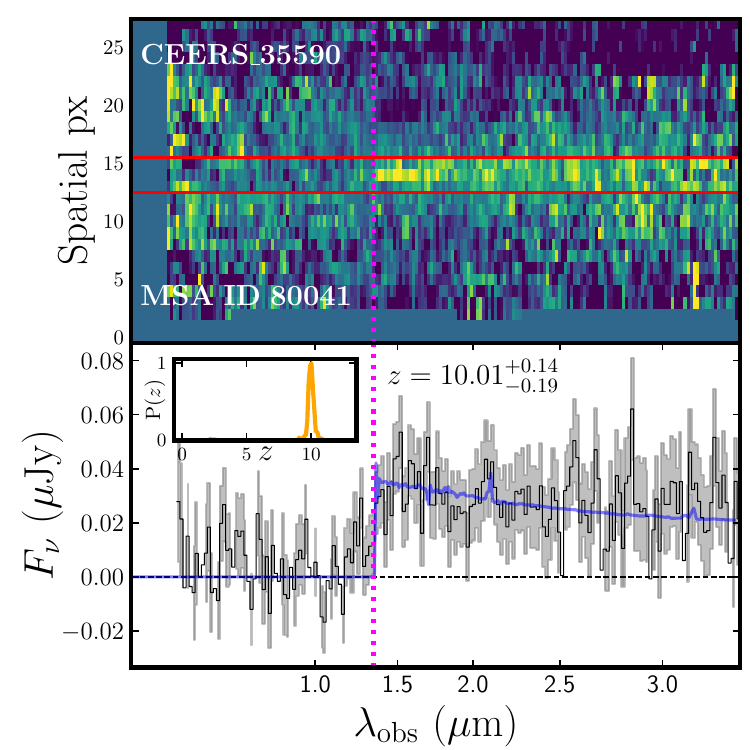}
\caption{2D (\textit{top}) and 1D (\textit{bottom}) spectra of CEERS\_99715 (\textit{left}) and CEERS\_35590 (\textit{right}). The wavelength axes have been limited to better visualize the Ly$\alpha$ break. The bottom panel inset shows the redshift probability distribution function. The best-fitting {\sc eazy} model is represented in blue, with a dotted vertical magenta line extending the derived Ly$\alpha$ break location to the 2D spectrum. The two horizontal red lines in the 2D spectrum indicate the 1D extraction window. The shaded grey area corresponds to the rescaled 1D flux errors (see Appendix~\ref{sec:appendix_noise}).}
\label{fig:break_zoom}
\end{figure*}

It is hard to measure precise redshifts of these two sources from breaks alone given the low $S/N$ of the spectra.
For CEERS\_99715, the best solutions obtained with the different Ly$\alpha$ break fitting methods are in good agreement with each other (see Table~\ref{tab:redshift}). As a result, as discussed in \S\ref{sec:redshift}, we adopt the result from the ``modified--{\sc eazy}'' method, which gives a Ly$\alpha$ break fit ($z=9.77^{+0.37}_{-0.29}$) as our best redshift for CEERS\_99715.
In the case of CEERS\_35590, the best solutions derived from the three Ly$\alpha$ break fitting methods described in \S\ref{sec:redshift} are more discrepant and inconsistent with the photometric redshift estimate. Moreover, if this source were actually at $z\approx8.4-9.2$ and considering the detection of strong UV continuum with a blue spectral slope (see \S\ref{sec:beta_slope}), we would expect \oiii4960,5008 and H$\beta$ emission lines to be unambiguously detected at $\lambda\approx4.6-5.1\ \mu$m, but these lines are not observed.
The best solution ($z = 10.01_{-0.19}^{+0.14}$) obtained when applying the ``modified--{\sc eazy}'' method to the full NIRSpec spectrum (not just the break region) is in much better agreement with both the photometric redshift, the location of the Ly$\alpha$ break and the absence of emission lines, which shift beyond the red wavelength limit of NIRSpec. We therefore adopt the result from the ``modified--{\sc eazy}'' fit to the full spectrum as our best redshift for this object.  This value is also consistent with a secondary peak in the redshift probability distribution function $P(z)$ derived by the MCMC1 break fitter, and is within the allowable range of $P(z)$ for MCMC2.
A closer look at the Ly$\alpha$ breaks of these two sources with their preferred break redshift fits is shown in Fig.~\ref{fig:break_zoom}. 

In any case, we caution about the uncertainties associated with the exact redshift estimation of $z>9.6$ galaxies under low-$S/N$ conditions. In particular, bright pixels due to un-removed artifacts such as detector flaws, cosmic rays, or ``snowballs'' can have a large impact on break fitting if such pixels occur below the break wavelength.  They can drive a break-fitting algorithm to lower redshift values in order to accomodate these few apparently significant pixels in the spectrum.

\subsubsection{The high redshift prism sample}

After removing the three undetected $z>8$ candidates, we end up with a sample of 4 NIRCam-selected $z\gtrsim8$ galaxies in CEERS epoch 3 observations. The two at $z<9$ present clear emission lines (see Fig.~\ref{fig:spectra_mosaic}) and therefore robust spectroscopic redshifts. For the two at $z\sim10$, their spectra strongly suggest a high-$z$ nature in agreement with their original photometric redshift estimates, but their exact redshifts are hard to determine given the absence of clear emission lines and the systematics in the redshift measured from the continuum break. Hence, the redshift derived for these two sources present larger uncertainties. However, the detected breaks and the absence of definitive emission lines in the NIRSpec spectral range support the inferred redshifts $z > 9.6$. The adopted spectroscopic redshifts for the NIRCam-selected $z>8$ candidates observed with NIRSpec in CEERS epoch 3 are summarized in Table~\ref{tab:fiducial_redshift}. Information about the three spectroscopically undetected targets is presented in Appendix~\ref{sec:appendix_CEERS_spec_compilation} along with all the other CEERS $z>8$ candidates in NIRSpec MSA observations without a robust redshift measurement \citep[see][]{ArrabalHaro2023, Fujimoto2023}.

\begin{table}
\caption{Adopted spectroscopic redshifts of CEERS epoch 3 NIRCam-selected $z>8$ candidates.}
\label{tab:fiducial_redshift}
\begin{center}
\begin{tabular}{lccc}
\hline\hline
Source ID  &  MSA ID  & $z_{\rm spec}$ & Method \\
(1)  &  (2)  &  (3) & (4) \\
\hline  
CEERS\_35590  &  80041  &  $10.01_{-0.19}^{+0.14}$ & {\sc eazy} full\\
CEERS\_99715  &  80026  &  $9.77_{-0.29}^{+0.37}$ & {\sc eazy} break\\
CEERS\_90671  &  80083  &  $8.638_{-0.001}^{+0.001}$ & \oiii \\
CEERS\_96937  &  80025  &  $7.651_{-0.001}^{+0.001}$ & \oiii \\
\hline \hline
\end{tabular}
\end{center}

\tablecomments{
(1) Source ID in the CEERS catalog (Finkelstein et al., in prep.).
(2) Source ID in the CEERS MSA observations.
(3) Best spectroscopic redshift. 
(4) Redshift estimation method (see \S\ref{sec:redshift}). The estimation based on \oiii5008 is prioritized in the cases where that line is detected. In the absence of high $S/N$ emission lines, the {\sc eazy} fit is to the Ly$\alpha$ break or to the complete spectrum when the break $S/N$ is too low for a robust estimation.
}
\end{table}

\subsection{Emission line measurements}
\label{sec:flux_EW}

For the two galaxies with $7.5 < z < 9$ we measure observed-frame equivalent widths (EWs) of the emission lines by fitting each 1D spectrum with a composite ``line + continuum'' model. Each line is modeled with a 1D Gaussian. The amplitude, spectral width, and center of the Gaussian are left as free parameters. The width ($\sigma$) is allowed to vary from 50 to 400 km s$^{-1}$ and the line center is allowed to vary over -600 to +600 km s$^{-1}$ from the systemic redshift of the galaxy. The continuum is modeled using a 1D polynomial (i.e., a line with a slope). To derive uncertainties on the fitted parameters, we adopt a Monte Carlo approach and create 1000 realizations of each 1D spectrum. The realizations are created by perturbing the observed spectrum by its error spectrum. For each realization, we re-fit the composite model. We estimate the uncertainties on the line fluxes and the level of the continuum from the distribution of the best-fit parameters of the suite of realizations.

The rest-frame EWs of the emission lines detected are listed in Table~\ref{tab:EWs}. There is no evidence of Ly$\alpha$ emission in any of the four sources presented in this work.

\begin{table}
\begin{center}
\caption{EW$_{0}$ for the lines detected in the prism spectra.}
\label{tab:EWs}

\begin{tabular}{lccccc}
\hline\hline
MSA ID  &  \oii  &  H$\gamma$  &  H$\beta$  &  \oiii   &  \oiii  \\
   &  3727  &   &    & 4960 & 5008 \\
   &    &   &  (\AA{})  &  &  \\
(1)  &  (2)  &  (3)  &  (4)  &  (5)  &  (6) \\
\hline  
  80083  & --- & $44^{+27}_{-19}$ & $56^{+31}_{-23}$ & $81^{+30}_{-18}$ & $287^{+68}_{-44}$ \\
 80025  & $57^{+30}_{-28}$ & --- &  $8^{+18}_{-8}$ & $51^{+17}_{-21}$ & $187^{+27}_{-23}$ \\
\hline \hline
\end{tabular}
\end{center}

\tablecomments{
(1) Source ID in the CEERS MSA observations.
(2)-(6) Rest-frame EW of different emission lines.
}
\end{table}

\subsection{SED fitting}
\label{sec:SED_fitting}

We carry out SED modeling using three independent fitting tools: {\sc bagpipes} \citep{Carnall2018, Carnall2019a}, {\sc cigale} \citep{Burgarella2005, Noll2009, Boquien2019} and Dense Basis \citep{Iyer2019}. For the sources with clear \oiii\ emission, the redshift is fixed to the adopted spectroscopic value (Table~\ref{tab:fiducial_redshift}) during the fitting. For the two $z\sim10$ galaxies, Gaussian redshift priors are allowed with a $\sigma$ similar to the larger uncertainty on the Ly$\alpha$ break redshifts. The SED fitting performed with {\sc cigale} and Dense Basis only make use of the NIRCam photometry, while the spectra is also included when using {\sc bagpipes}. In all cases, we assume a \cite{Chabrier2003} Initial Mass Function (IMF).  

All sources are fit with \textsc{bagpipes} assuming a \cite{Calzetti2000} dust law. We fit with BPASS v2.2.1 \citep{eldridge17} stellar templates over a range of ionization parameter $\log U \in [-4,-1]$ and total (stellar and gas-phase) metallicity $Z \in [-3,1]$, a flexible star formation history (SFH) represented by a Gaussian Mixture Model \citep[GMM, ][]{Iyer2019} and a log-normal prior on star formation rate (SFR) over the range $\log(\mathrm{SFR}) \in [-2,3]$. We first scale the individual spectra to match the observed photometry for each source, then fit the source photometry and spectra simultaneously, assuming a $\chi^2$-likelihood function, to infer galaxy properties from the posterior distributions. When fitting CEERS\_35590 we include the MIRI F560W and F770W photometry (\S\ref{sec:MIRI_data}). To determine if the inclusion of the MIRI photometry has any impact on the inferred stellar mass and SFR for CEERS\_35590, we fit the CEERS\_35590 photometry and spectrum a second time, but exclude the MIRI photometry. In both cases, we infer a stellar mass of $\log(M_*/M_\odot) = 9.1 \pm 0.1$ and $\log(\mathrm{SFR}) = 9 \pm 2$ and therefore adopt the fit that includes the MIRI photometry as the best fitting Bagpipes model.

Similarly, sources are fit with the Dense Basis SED fitting code using flexible nonparametric SFHs \citep{Iyer2019}. 
For this work, we define 3 ``shape'' parameters that describe the SFH: $t_{25}, t_{50},$ and $t_{75}$ (requiring the recovered SFH of the galaxy to form ``$x$'' fraction of its total mass by time $t_x$). We impose a uniform (flat) prior on the specific star formation rate (sSFR) with limits on the SFR (SFR$/M_\odot\  \textrm{yr}^{-1} \in [10^{-2}, 10^{3}]$), an exponential prior on the dust attenuation over a wide range of values ($A_V \in [0, 4]$), and a uniform (in log-space) prior on the metallicity ($Z/Z_\odot \in [0.01, 2.0]$).

Finally, we fit the objects with {\sc cigale}. We select a delayed SFH with SFR $\propto t \times \exp(-t/\tau)$. An additional final burst of star formation ($k\times \exp(-t/\tau_{\mathrm{burst}})$) is introduced if it provides a better fit. The age of the main stellar population age$_{\mathrm{main}}$ is allowed to vary from 2 Myr to 1 Gyr while age$_{\mathrm{burst}}$ is set to 1 Myr and the burst fraction ($f_{\mathrm{burst}} \equiv M_{\star,\mathrm{burst}} / M_{\star,\mathrm{tot}}$) is free in the range $f_{\mathrm{burst}} \in [0, 0.5]$. BPASS v2.2.1 stellar templates with a fixed $Z = 0.008$ metallicity are used. 
Dust attenuation following \cite{Calzetti2000} is applied to the stellar continuum, while nebular emission (continuum and lines) is attenuated with a screen model and a Small Magellanic Cloud (SMC) extinction curve \citep[][]{Pei1992}.

\begin{figure*}
\centering
\includegraphics[width=\textwidth]{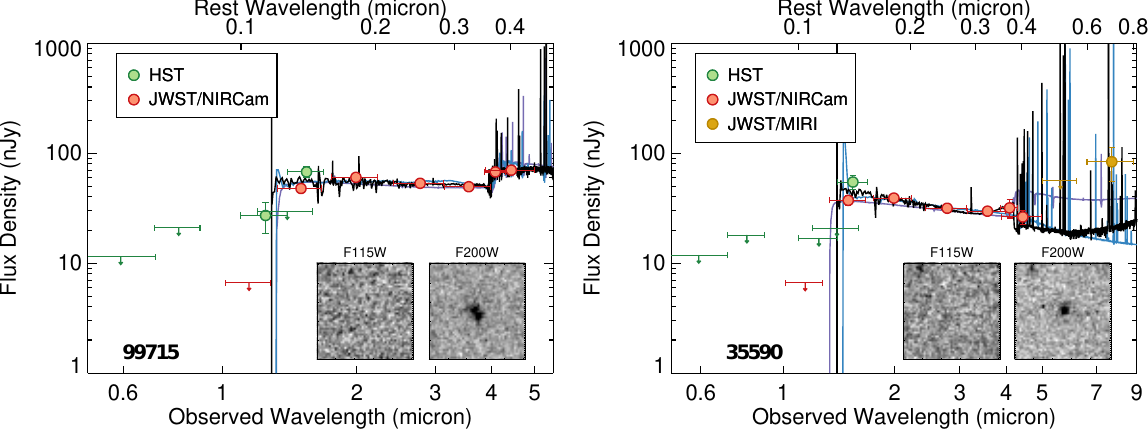}\\
\includegraphics[width=\textwidth]{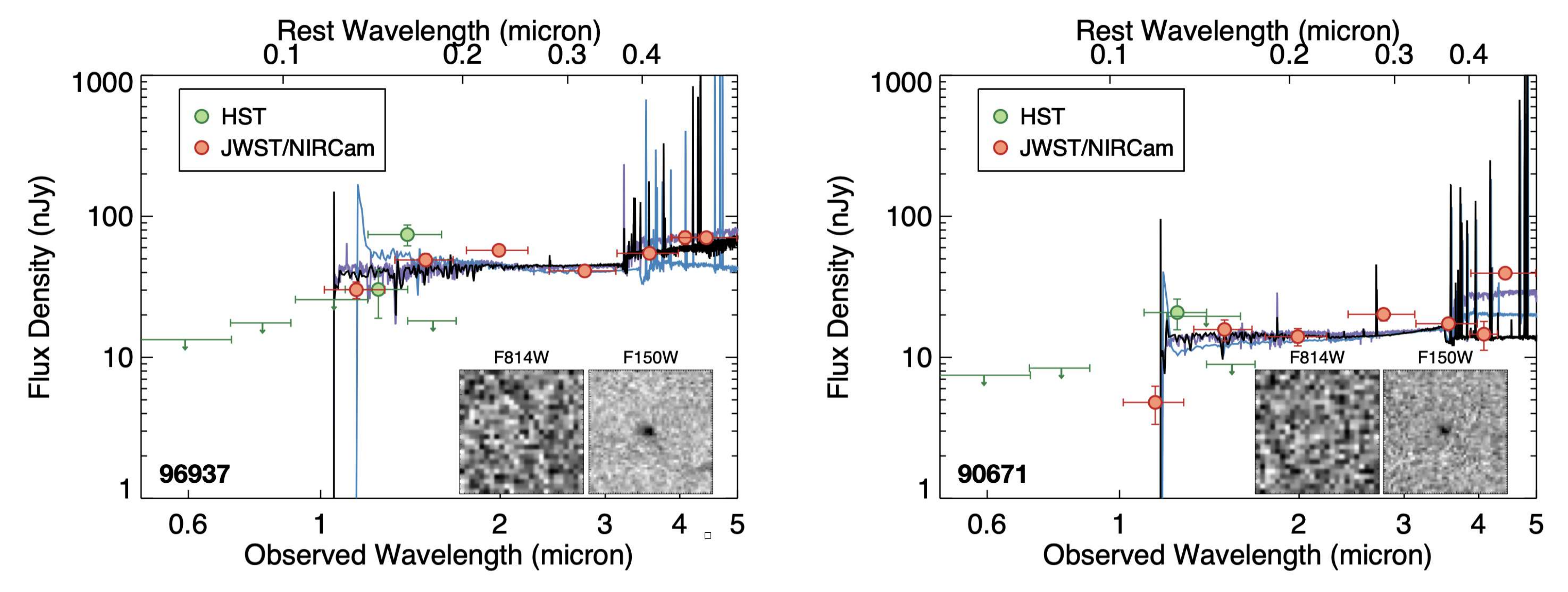}\\

\caption{SED of the four spectroscopically confirmed sources. The measured photometry in ACS F606W and F841W filters is represented in green; NIRCam F115W, F150W, F200W, F277W, F356W, F410M and F444W, in red; and MIRI F560W and F770W (for CEERS\_35590 only), in orange. The spectra represents the best-fitting stellar population models, with black, blue and purple representing {\sc cigale}, {\sc Dense Basis} and {\sc Bagpipes}, respectively. The inset stamps are $1.5''\times1.5''$ cutouts of some combination of the NIRCam F115W, F150W, F200W and ACS F814W images highlighting the emission drop corresponding to the Ly$\alpha$ break.}
\label{fig:SED_fit_and_stamps}
\end{figure*}

The main physical parameters derived for our four objects with each SED fitting tool are summarized in Table~\ref{tab:physical_properties}, and the SEDs and best-fitting models of the two $z\sim10$ objects are presented in Fig.~\ref{fig:SED_fit_and_stamps}.

The MIRI photometric measurements for CEERS\_35590 have low $S/N = 1.7$ and 2.9 for F560W and F770W, respectively, but both measurements (especially F770W) are well above a simple extrapolation of the NIRCam photometry to $\lambda > 5\,\mu$m (see Fig.~\ref{fig:SED_fit_and_stamps}). They are consistent with the presence of strong [OIII]+H$\beta$ emission in F560W and H$\alpha$+[NII] in F770W.  

Furthermore, an additional fit of the two $z\sim10$ sources is carried out with {\sc bagpipes} making use of the photometry and spectra in the same way described above but without imposing any constraints on the redshift. The best redshift solutions obtained in this case are $z=9.72^{+0.06}_{-0.04}$ for CEERS\_99715 and $z=9.97^{+0.10}_{-0.07}$ for CEERS\_35590, in very good agreement with our adopted fiducial values (see Table~\ref{tab:fiducial_redshift}).

\begin{table}
\begin{center}
\caption{Stellar mass, SFR, sSFR and dust attenuation.}
\label{tab:physical_properties}
\footnotesize
\begin{tabular}{llcccc}
\hline\hline
Code & MSA  &  $\log (M_{\star}/M_{\odot})$  &  SFR  &  sSFR  &  $A_V$ \\
   & ID &  &  {\footnotesize ($M_{\odot}\ \mathrm{yr}^{-1}$)} & $\log (\mathrm{yr^{-1}})$  & (mag) \\
(1)  &  (2)  &  (3)  &  (4) & (5) & (6) \\
\hline  
 B  & 80041  & $9.1^{+0.1}_{-0.1}$ &  $9^{+2}_{-2}$   &  $-8.2^{+0.1}_{-0.2}$ &  $0.1^{+0.1}_{-0.1}$ \\
                 & 80026  & $9.5^{+0.1}_{-0.1}$ &  $6^{+4}_{-2}$   &  $-8.7^{+0.3}_{-0.3}$  &  $0.1^{+0.1}_{-0.1}$ \\
                 & 80083  & $8.8^{+0.1}_{-0.1}$ &  $3^{+2}_{-2}$   &  $-8.3^{+0.2}_{-0.4}$ &  $0.4^{+0.2}_{-0.2}$ \\
                 & 80025  & $9.3^{+0.1}_{-0.1}$ &  $6^{+1}_{-1}$   &  $-8.5^{+0.2}_{-0.2}$  &  $0.4^{+0.1}_{-0.1}$ \\
\hline
C  & 80041  & $8.2^{+0.2}_{-0.4}$ &  $6^{+4}_{-4}$   & $-7.4^{+0.6}_{-0.7}$ &  $0.1^{+0.1}_{-0.1}$ \\
              & 80026  & $9.0^{+0.2}_{-0.3}$ &  $11^{+15}_{-11}$   & $-7.9^{+0.6}_{-1.2}$ &  $0.4^{+0.3}_{-0.3}$ \\
              & 80083  & $8.2^{+0.2}_{-0.4}$ &  $8^{+7}_{-7}$   & $-7.3^{+0.7}_{-1.1}$ &  $0.9^{+0.4}_{-0.4}$ \\
              & 80025  & $8.8^{+0.1}_{-0.2}$ &  $8^{+10}_{-8}$   & $-7.9^{+0.6}_{-1.0}$ &  $0.6^{+0.3}_{-0.3}$ \\
\hline  
DB   & 80041  & $8.7^{+0.4}_{-0.3}$ &  $3^{+21}_{-1}$  & $-8.2^{+1.2}_{-0.6}$ &  $0.1^{+0.2}_{-0.1}$ \\
   & 80026  & $9.2^{+0.2}_{-0.4}$ &  $9^{+4}_{-4}$  & $-8.3^{+0.6}_{-0.5}$ &  $0.1^{+0.2}_{-0.1}$ \\
        & 80083  & $8.8^{+0.0}_{-1.0}$ &  $1^{+1}_{-1}$  & $-8.7^{+1.0}_{-0.9}$ & 
 $0.5^{+0.3}_{-0.3}$ \\
        & 80025  & $8.6^{+0.1}_{-0.1}$ &  $0.4^{+11}_{-0.3}$  & $-9.4^{+2.0}_{-0.4}$ & 
 $0.0^{+0.1}_{-0.0}$ \\
\hline \hline
\end{tabular}
\end{center}

\tablecomments{
(1) SED-fitting code used to derive the physical properties (B: \textsc{Bagpipes}, C: \textsc{cigale}, DB: Dense Basis).
(2) Source ID in the CEERS MSA observations.
(3) Stellar mass.
(4) SFR averaged over the last 100 Myr.
(5) Specific SFR.
(6) Stellar dust extinction.
}
\end{table}

\subsection{UV spectral slope}
\label{sec:beta_slope}

We measure the UV spectral slope $\beta$ from the photometry, the spectra and the best SED-fitting models using the adopted redshifts in Table~\ref{tab:fiducial_redshift}.

For the photometry, we fit the WFC3 and NIRCam SED  with a power law ($f_\lambda \propto \lambda^{\beta}$) between 1500--3000~\AA{} rest-frame \citep[see][]{Calzetti1994}.
Using the \textsc{emcee} software \citep{foreman-mackey13}, we measure the posterior distribution on $\beta$ and obtain the median and 68\% central width from the posterior. 
The same process is employed to measure the  $\beta$ slope directly from the prism spectra. We note here that the comparison of the spectrum flux density with the photometry of the four objects here presented reveals small flux discrepancies (by a factor $<2$), but these deviations are similar along the prism wavelengths without a particular trend, so we expect the spectroscopic $\beta$ not to be significantly affected by the absence of a precise slit loss correction.
The model-derived $\beta$ is retrieved from the best {\sc bagpipes} models resulting from the SED fitting employing both photometry and spectra (see \S\ref{sec:SED_fitting}). 

The absolute UV magnitude $M_{\mathrm{UV}}$ at 1500~\AA{} is estimated following the methodology used in \cite{Finkelstein2015}. Table~\ref{tab:beta_slope} presents the $\beta$ values obtained from the three different estimations as well as the $M_{\mathrm{UV}}$ values for our four galaxies.

\begin{table}
\begin{center}
\caption{$M_{\mathrm{UV}}$ and independent UV $\beta$ slopes.}
\label{tab:beta_slope}

\begin{tabular}{lcccc}
\hline\hline
MSA ID  & $M_{\mathrm{UV}}$  &  $\beta_{\mathrm{phot}}$  &  $\beta_{\mathrm{model}}$  &  $\beta_{\mathrm{spec}}$ \\
(1)  &  (2)  &  (3)  &  (4)  &  (5) \\
\hline  
 80041  &  $-20.1^{0.1}_{0.1}$  & $-2.19_{-0.56}^{+0.92}$ & $-2.33^{+0.06}_{-0.06}$ & $-1.93_{-0.52}^{+0.55}$ \\
 80026  &  $-20.5^{0.1}_{0.1}$  & $-2.16_{-0.52}^{+0.78}$ & $-2.15^{+0.05}_{-0.05}$ & $-1.87_{-0.52}^{+0.54}$ \\
80083  &  $-18.7^{0.1}_{0.1}$  & $-1.45_{-0.71}^{+0.74}$ & $-1.90^{+0.17}_{-0.20}$ & $-1.57_{-0.46}^{+0.37}$ \\
 80025  &  $-20.0^{0.2}_{0.1}$  & $-2.32_{-0.48}^{+0.93}$ & $-1.59^{+0.08}_{-0.08}$ & $-1.86_{-0.56}^{+0.57}$ \\
\hline \hline
\end{tabular}
\end{center}

\tablecomments{
(1) Source ID in the CEERS MSA observations.
(2) Absolute UV magnitude measured at 1500~\AA{} rest-frame.
(3) Photometric UV slope.
(4) UV slope derived from the {\sc bagpipes} models best-fitting photometry and spectra simultaneously.
(5) Spectroscopic UV slope.
}
\end{table}

\subsection{Morphology}

We use \texttt{Galfit} \citep{peng02,peng10}, a least-squares fitting algorithm that finds the optimum S\'ersic fit to a galaxy’s light profile, to measure the size and morphology of the four spectrosopically confirmed $z>8$ galaxies listed in Table \ref{tab:redshift}. As input, we create 100$\times$100 pixel cutouts of the F200W and F277W images (with a 0.03$''$ pixel scale), the error array to use as the input sigma image, and the source segmentation map. We use empirically derived PSFs based on stacked stars from the image. As an initial guess of the parameters we use the source location, magnitude, size, position angle, and axis ratios from the \texttt{Source Extractor} \citep{Bertin1996} catalog. We then run \texttt{Galfit}, allowing the Sersic index to vary between 0.01 and 8, the magnitude of the galaxy between 0 and 45, and the size (R$_e$) between 0.3 and 200 pixels and allowing the PSF to be oversampled by a factor of 9. To make sure the fits were reasonable, we then visually inspect the best-fit model and image residual.

\begin{figure}
\includegraphics[width=\columnwidth]{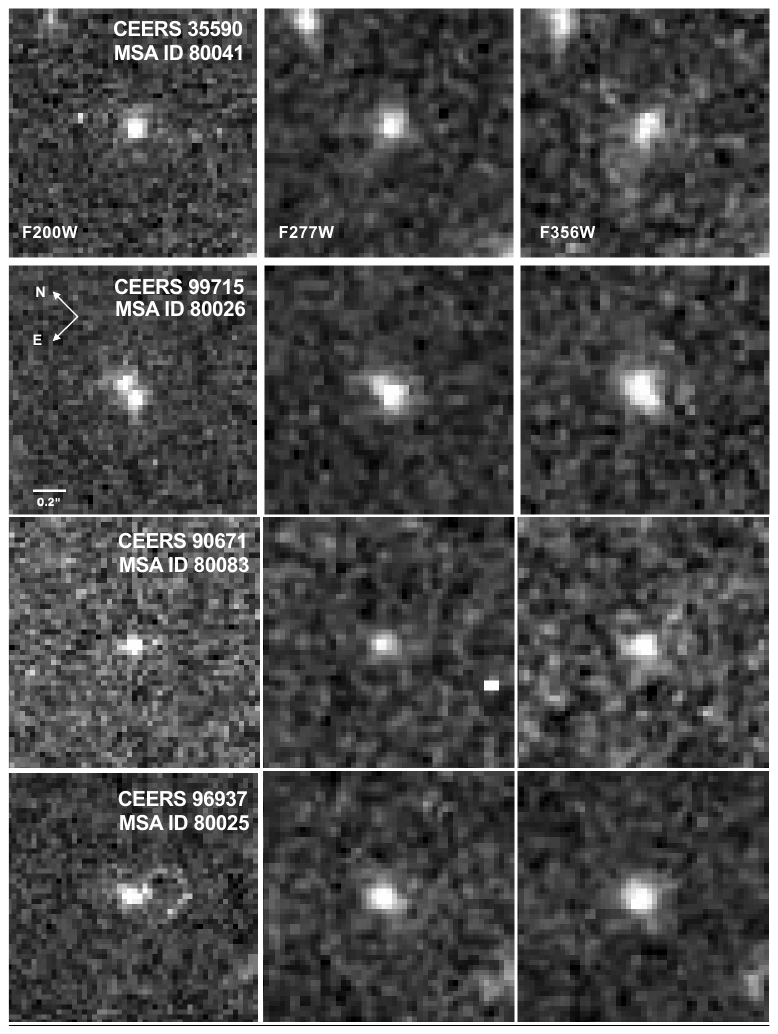}
\caption{Cutout images of the four $z>8$ sources in F200W, F277W, and F356W, highlighting the morphological structure visible at different wavelengths. Each cutout is  1.5$''\times1.5''$.}
\label{fig:morph_cutouts}
\end{figure}

At F277W, CEERS\_35590 is well-resolved and fit with a single S\'ersic component. We measure a half-light radius of $3.31 \pm 0.18$ pixels (0.42 $\pm$0.02 kpc) with a S\'ersic index of $n=0.75$. At F200W, \texttt{Galfit} hits the constraint limits, indicating that the source is not well-fit. Visually, CEERS\_35590 is very compact at F200W and more extended at longer wavelengths with what appears to be extended faint emission at F356W and F444W to the east of the galaxy (see Fig.~\ref{fig:morph_cutouts}).

CEERS\_99715 is well-resolved at both F200W and F277W and we are able to fit both filters with a single S\'ersic component. The source has different visual morphologies in F200W and F277W, with two components visible at shorter wavelengths (see Fig.~\ref{fig:morph_cutouts}). At F200W we measure a half-light radius of $4.53 \pm 0.19$ pixels (0.58 $\pm$0.02 kpc) with $n=0.60$ and at F277W we measure a half-light radius of $4.86 \pm 0.17$ pixels (0.62 $\pm$0.02 kpc) with $n=0.77$.

CEERS\_90671 is unresolved at F200W and marginally resolved at F277W with a half-light radius of $2.86 \pm 0.22$ pixels (0.36 $\pm$0.03 kpc) and $n=0.68$. CEERS\_96937 is well-resolved at F277W, with a half-light radius of $16.50 \pm 2.64$ pixels (2.09$\pm$0.34 kpc) and $n=2.53$.  At F200W, the fit is likely affected by what appears to be a snowball artifact. We measure a half-light radius of $4.07 \pm 0.76$ pixels (0.52 $\pm$0.10 kpc) and $n=0.61$.

\section{Results and discussion}
\label{sec:results}

\subsection{Physical properties}

The main physical properties derived through SED fitting for the four spectroscopically confirmed galaxies are summarized in Table~\ref{tab:physical_properties}. Clear differences can be appreciated in the values derived from each SED fitting code for some sources, as expected when the SFH cannot be well constrained at high-$z$ \citep[see, e.g.,][]{Carnall2019b, Leja2019, Pacifici2023}. 
These differences illustrate the difficulty of determining stellar population parameters even at this very early cosmic epoch when galaxy ages are necessarily young.
In particular, we obtain systematically higher $A_V$ (although compatible within the errors) from the {\sc cigale} fitting, as well as lower stellar masses.

In this section we focus on the physical properties of the two $z\sim10$ sources presented in this work, refering to \cite{Fujimoto2023} and Arrabal Haro et al. (in prep.) for a more general discussion of the complete CEERS $z=8-9$ spectroscopic sample. 
The confirmation of these two objects raises to six the number of spectroscopically confirmed luminous sources with $M_{\mathrm{UV}}<-20$ at $z \gtrsim 10$, the others being GN-z11 \citep{Oesch2016, Bunker2023, Tacchella2023}, Maisie's galaxy \citep{Finkelstein2022b, ArrabalHaro2023}, CEERS\_11384 \citep{ArrabalHaro2023, Harikane2023b} and M0647-JD \citep{Harikane2023b, Hsiao2023}. This suggests that these luminous sources appeared more often than expected early in the history of the universe ($<500$~Myr).

The stellar masses derived are among the largest of those reported in \cite{ArrabalHaro2023}, \cite{Bunker2023}, \cite{Curtis-Lake2023} and \cite{Roberts-Borsani2022} for spectroscopically confirmed $z\gtrsim10$ galaxies. While we cannot robustly constrain the stellar mass of CEERS\_35590 ($\log(M_{\star}/M_{\odot})\simeq8.2-9.1$), we find a relatively high $\log(M_{\star}/M_{\odot})\simeq9.0-9.5$ for CEERS\_99715 (see Table~\ref{tab:physical_properties}), similar to those of GN-z11 and GS-z11-0 \citep{Curtis-Lake2023, Robertson2023}. Considering the still limited spectroscopic follow up of $z>10$ candidates to date, the confirmation of already three objects with $\log(M_{\star}/M_{\odot})\geq9$ suggests that the abundance of such evolved systems is higher than predicted \citep[e.g.,][]{Finkelstein2022b, Finkelstein2023, Harikane2023a}, although it is important to keep in mind that these are the easiest sources to detect among the $z>10$ candidates. Nevertheless, the Santa Cruz semi-analytic model \citep[SC SAM; e.g.][]{Somerville2015, Somerville2021} predicts $0.3^{+2}_{-0}$ galaxies with $\log(M_{\star}/M_{\odot})\sim9$ at $9.48 < z < 10.15$ (defined to account for our $z$ errors) in the whole NIRCam imaging area covered by CEERS \citep[see][]{Yung2019b}. Although confirming a single galaxy under those conditions (and a tentative one) is still consistent within the uncertainties of the prediction, it is likely that the actual number of galaxies at those redshifts and stellar mass is larger if we consider that the MSA observations presented here are a lower limit for any kind of number density estimation. Similarly, the SC SAM predicts $1.1^{+2.7}_{-0.4}$ sources with $M_{\mathrm{UV}}\sim-20.5$ at the same redshifts \citep{Yung2019a}, again still compatible with the lower limit of two objects here presented, but likely underpredicting the actual number when we consider the small fraction of spectroscopic follow-up of the CEERS field. 
Predictions from the SIMBA-EoR hydrodynamic simulation \citep{Dave2019} indicate that we should see $\sim1$ galaxy at $8<z<10$ with SFR $>$ 3~$M_{\odot}\ \mathrm{yr^{-1}}$, highly inconsistent with the lower limit of 4-6 (depending on assumed SFHs during SED fitting) already reported in \cite{Fujimoto2023} and this work. Reducing stellar feedback in the simulations \citep[see, e.g.,][]{Dekel2023, Yung2023} raises the number of expected galaxies in the SIMBA-EoR simulation to $\sim7$, again leaving not much room before the observations are in tension with even the no-feedback simulations.

The low dust attenuation measured for the two $z\sim10$ sources $A_{V}\simeq0.1^{+0.2}_{-0.1}$ is in very good agreement with similarly low values ($A_{V}\leq0.3$) reported for all the confirmed $z\gtrsim10$ objects in the literature \citep{ArrabalHaro2023, Hsiao2023, Robertson2023, Tacchella2023}, often compatible with zero attenuation within the errors. This is interesting as a hypothetical scenario with negligible dust attenuation could be one of the factors that would increase the predicted numbers of bright galaxies in the early universe in some theoretical models \citep{Ferrara2023}. Consistent with the low dust extinction measured, the spectra of the two $z\sim10$ galaxies, present blue continua, although not extremely blue, with $-2.3\lesssim\beta\lesssim-1.9$ for the independent estimations (see Table~\ref{tab:beta_slope}).

Finally, we measure compact sizes for the two $z\sim10$ objects, with half-light radius $\sim0.4-0.6$ kpc, consistent with sizes previously measured for objects at the EoR \citep[see, e.g.,][]{Mascia2023, Treu2023}. Interestingly, we see some variation of the global morphology of both sources with wavelength, contrary to the observed behavior in \cite{Treu2023}. In particular, the two resolved components observed for CEERS\_99715 at the shorter wavelengths present an interesting  case study for future resolved SED fitting analysis that could help distinguish whether this is a merger of two distinct galaxies, or if they are two ``clumps'' within one galaxy with spatially distinct regions of star formation or separated by a dust lane. If this is an ongoing merger it may illustrate the processes that lead to early galaxies as massive as CEERS\_99715 ($\log(M_{\star}/M_{\odot})\simeq9.3^{+0.2}_{-0.3}$; average from the three independent estimations), as discussed in \cite{Boyett2023}.

\subsection{Robustness of photometric redshifts}

The CEERS epoch 3 NIRSpec observations presented here complete the Cycle 1 spectroscopic follow up of $z > 8$ galaxy candidates in the CEERS field. Here we consider the CEERS NIRSpec observations together with one additional NIRSpec field observed in DD time program \#2750 (PI: Arrabal Haro), which also targeted CEERS NIRCam-selected high redshift candidates using longer exposure times (18387~s, compared to 3107~s for CEERS).  Together, these NIRSpec MSA observations targeted 32 candidates that meet the $z > 8$ photometric redshift selection criteria of F23. Merging the spectroscopic measurements from \cite{ArrabalHaro2023}, \cite{Fujimoto2023}, \cite{Tang2023} and from the present work, 24 out of the 32 targets have robust spectroscopic redshifts. 

Eight targets meeting the F23 $z_\mathrm{phot}$ criteria do not have securely measured spectroscopic redshifts.
Seven of these 8 objects are faint ($m_{\mathrm{F277W}}\gtrsim28$) and have  $z_{\mathrm{phot}}>10$, where strong emission lines would not be expected at $\lambda < 5.3\,\mu$m and where continuum $S/N$ is low with the exposure times employed for the CEERS NIRSpec observations. None of these 8  candidates shows emission lines that would be expected if they were at $z \lesssim 9.6$ \citep[see also][]{Fujimoto2023}. Therefore, it seems likely that these galaxies have $z > 9.6$, despite the lack of secure spectroscopic confirmation.

Several other authors have also identified high-redshift galaxy candidates from CEERS NIRCam data using different selection criteria. Some of those candidates have also been observed with NIRSpec, including two targeted but undetected sources from  \cite{Donnan2023} and \cite{Whitler2023}, plus a third $z\sim8$ candidate from \cite{Labbe2023} with spectroscopic $z = 5.623$ \citep{Kocevski2023}.  In total, 35 CEERS $z>8$ candidates from the literature have been followed up with NIRSpec MSA observations, 25 of them with robust spectroscopic redshift measurements.

Appendix~\ref{sec:appendix_CEERS_spec_compilation} presents a complete census of spectroscopically observed $z_{\mathrm{phot}}>8$ candidates in CEERS from \cite{Bouwens2023}, \cite{Donnan2023}, \cite{Endsley2022}, \cite{Finkelstein2023}, \cite{Harikane2023a}, \cite{Labbe2023} and \cite{Whitler2023}.
For completeness, we include one object with lower photometric redshift ($z_{\mathrm{phot}} = 6.45$) that is unambiguously confirmed to have spectroscopic $z = 8.175$ (MSA ID 1149; \citealt{Heintz2023}, \citealt{Sanders2023}, \citealt{Tang2023}).

\begin{figure*}
\centering
\includegraphics[width=\textwidth]{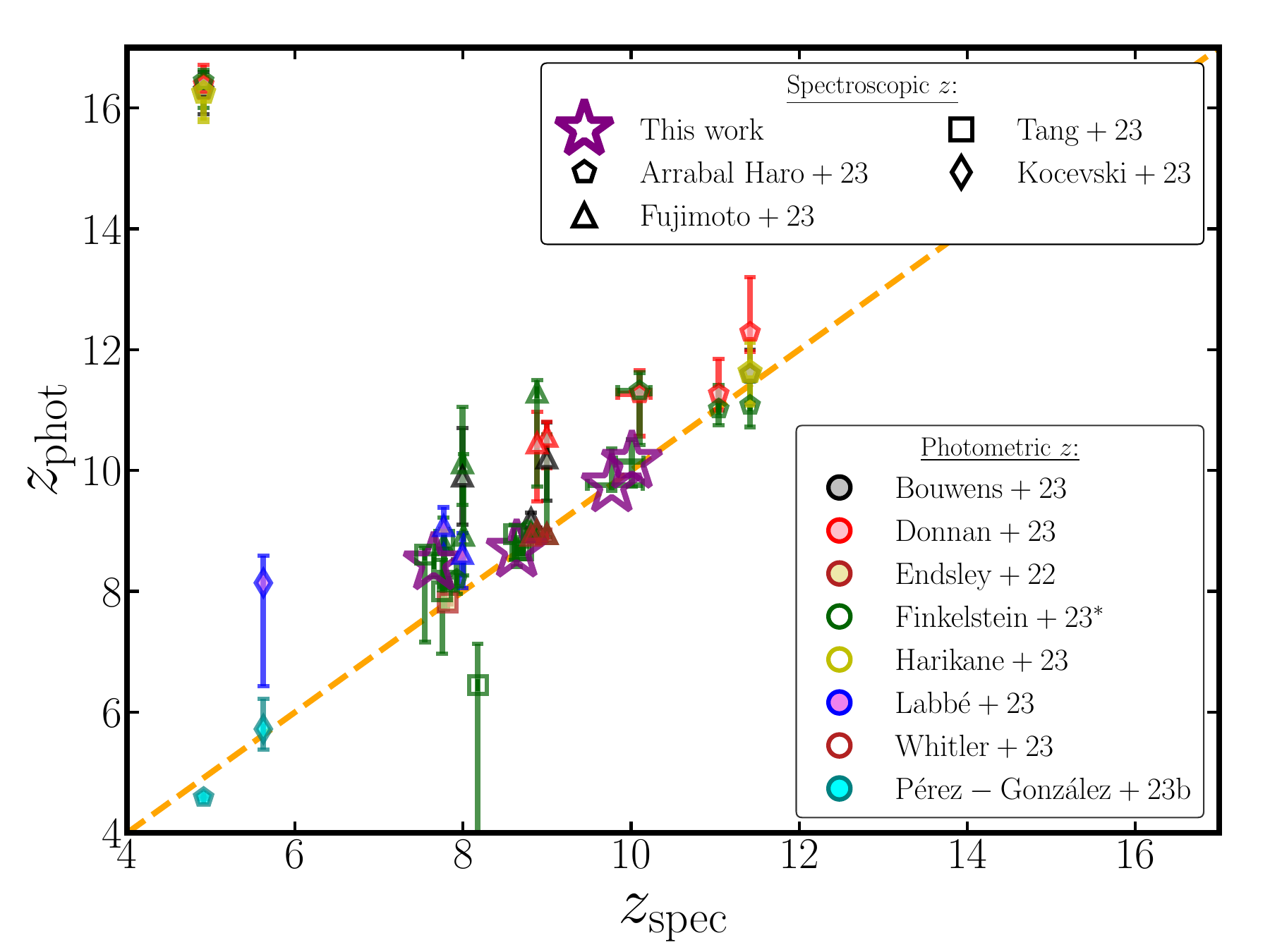}
\caption{Comparison of photometric and spectroscopic redshifts for $z_{\mathrm{phot}}>8$ candidates in CEERS from \cite{Bouwens2023}, \cite{Donnan2023}, \cite{Endsley2022}, \cite{Finkelstein2023}, \cite{Harikane2023a}, \cite{Labbe2023} and \cite{Whitler2023} with secure spectroscopic redshifts from the complete NIRSpec data in the EGS field. $^*$The green markers correspond to galaxies in \cite{Finkelstein2023} or selected using the selection criteria employed in that work (see \S\ref{sec:NIRCam_data}). For completeness, photometric estimations from the $2\lesssim z \lesssim 7$ sample in \cite{Perez-Gonzalez2023a} for the two lower-$z$ interlopers are also included. The spectroscopic redshifts are remeasured here for galaxies with emission lines spectroscopically reported in \cite{ArrabalHaro2023}, \cite{Fujimoto2023}, \cite{Kocevski2023} and \cite{Tang2023}. The four galaxies presented in this work for the first time are highlighted with purple stars. The one-to-one relation is shown as a dashed orange line. Note that several of the points represented in this figure correspond to independent photometric estimations of the same galaxies (therefore at the same $z_\mathrm{spec}$ value). A complete compilation of the redshifts here presented can be found in Appendix~\ref{sec:appendix_CEERS_spec_compilation}.}
\label{fig:z_spec_vs_z_phot}
\end{figure*}

A comparison of the spectroscopic and photometric redshifts for these high-redshift candidates is presented in Fig.~\ref{fig:z_spec_vs_z_phot}. This includes objects spectroscopically confirmed in \cite{ArrabalHaro2023}, \cite{Fujimoto2023}, \cite{Kocevski2023} and \cite{Tang2023}, whose redshifts have been remeasured for consistency, following the methodology described for emission lines redshifts in \S\ref{sec:redshift}. The good redshift agreement overall suggests that photometric high-$z$ selection methods perform quite well at identifying these sources. Out of the 25 galaxies with robust spectroscopy, only 2 (8\%) are demonstrated to be lower-redshift interlopers. 
The first of these sources is CEERS\_24015, originally selected as a massive $z\sim8$ candidate in \cite{Labbe2023} and later confirmed as a low-luminosity active galactic nucleus (AGN) at $z = 5.623$ by \cite{Kocevski2023}. The other interloper is CEERS\_13256, originally included in several high-$z$ selections \citep{Bouwens2023, Donnan2023, Finkelstein2023, Harikane2023a} and recently confirmed at $z=4.91$ by \cite{ArrabalHaro2023}. It is worth noting that both of these interlopers were also assigned lower photometric redshifts (consistent with their spectroscopic values) by other authors (see, e.g., \citealt{Perez-Gonzalez2023a} and discussion in \citealt{Naidu2022b, Zavala2023}). This demonstrates that even the few confirmed interlopers can be photometrically identified under certain selection criteria.

It is important to note that the high-$z$ photometric samples discussed here are not merely defined by the most likely redshift values from template-fitting; they also meet additional color and $\chi^2$ criteria that are used to define a high-fidelity sample \citep[see, e.g., detailed discussion in][]{Finkelstein2023}. In Appendix~\ref{sec:appendix_possible_interlopers} we show an illustrative example of a galaxy whose photometric redshift $P(z)$ peaks at $z_\mathrm{best} > 8$, but which does not meet other high-$z$ selection criteria of \cite{Finkelstein2023}, and for which NIRSpec measures $z = 5.271$. Objects like this could appear more frequently in high-$z$ photometric samples with more relaxed selection criteria. Therefore, the high confirmation rate (92\% among the detected targets) we find in this work can only be extrapolated to other high-$z$ samples based on selection criteria similar to the ones employed in the studies analyzed here and should not be taken for granted for less rigorously selected samples. 

In spite of the high ratio of confirmed high-$z$ sources, we observe a trend where the photo-$z$ is often slightly overestimated. Considering a mean photometric redshift for each individual object among their estimations in different studies, and after removing the low-$z$ interlopers, we compute an average deviation $\langle\Delta z\equiv z_{\mathrm{phot}}-z_{\mathrm{spec}}\rangle = 0.45\pm0.11$. 
This could, in part, be due to a confirmation bias that would cause a larger number of objects at lower redshifts to be well detected spectroscopically as they would be typically brighter. This effect would be particularly relevant at $z\gtrsim 9.6$, where strong rest-frame optical emission lines start to redshift beyond the red wavelength limit of the NIRSpec instrument, making lower spectroscopic redshifts easier to confirm via the presence of H$\beta$ and \oiii\ in the spectrum. However, the same trend is observed at $z_{\mathrm{spec}}=8-9$ where strong lines are detected with NIRSpec but where several galaxies have higher photometric redshifts. This suggests a deeper reason for this deviation.

It is also possible that there are physical differences in the true continuum shapes of very high-$z$ galaxy spectra which are not accounted for in the templates used for photometric redshift estimation.
In any case, these results suggest that our current $z>8$ selection criteria are good, but at the same time we might need to modify our galaxy templates to better resemble the SEDs of very high-$z$ galaxies and further refine our photo-$z$ estimations. For this purpose, deep continuum spectroscopy of a variety of early galaxies will be necessary to build the proper database of templates.

All things considered, the low fraction of interlopers reported in this work for $z>8$ candidates in CEERS (8\%) is insufficient to account for the high number densities reported in several studies \citep{Donnan2023, Finkelstein2022b, Finkelstein2023, Harikane2023a} with respect to the theoretical predictions. Therefore, these spectroscopic observations, while incomplete, reinforce the photometric evidence for an abundant population of comparatively luminous galaxies at $z>8$, extending at least to $z \simeq 10$ and perhaps beyond.  For galaxies at $z > 9.6$, where the strongest optical rest frame emission lines shift out of the NIRSpec spectral range, the 3107~s exposure times of CEERS NIRSpec observations limit redshift confirmation to brighter sources. Deeper surveys are needed to achieve higher spectroscopic completeness and to probe fainter down the galaxy luminosity function.

\section{Summary}
\label{sec:summary}

We present spectroscopic follow up of seven $z>8$ candidates in CEERS epoch 3 NIRSpec MSA observations. Four out of the seven targets are spectroscopically detected.
Two of them present clear rest-optical emission lines that result in unambiguous redshifts $z=8.638$ and $z=7.651$.
The other two detected sources present clear Ly$\alpha$ continuum breaks, but not emission lines. We determine their redshifts by fitting the continuum break through different methods, adopting fiducial values of $z=9.77_{-0.29}^{+0.37}$ and $z=10.01_{-0.19}^{+0.14}$. 
We show that spectroscopic redshifts based only on fitting the Ly$\alpha$ break are effective to confirm the high-$z$ nature of these objects, but have relatively large uncertainties, especially when working with low $S/N$ breaks, due to the low spectral resolution of the NIRSpec prism at the bluest wavelengths.
The three undetected targets do not show emission lines that identify them as lower-$z$ interlopers, therefore they remain consistent with $z\gtrsim9.6$ scenarios despite not having robust redshift determinations.

For the two $z\sim10$ sources, we measure relatively high luminosities ($M_{\mathrm{UV}}<-20$), blue UV slopes ($-2.3\lesssim\beta\lesssim-1.9$) and low dust extinction ($A_{V}\simeq0.15^{+0.3}_{-0.1}$). The object CEERS\_99715, in particular, presents a high stellar mass ($\log(M_{\star}/M_{\odot})=9.0-9.5$). Additionally, its morphological analysis reveals two differentiated substructures at the bluer wavelengths (NIRCam/F200W) that might be hints of a minor merger which could help to understand the way objects like this built a relatively large stellar mass only $\sim485$ Myr after the Big Bang.

The spectroscopic results presented here are combined with all the available $z>8$ spectroscopy from previous works for a complete census of NIRSpec MSA spectroscopy in the EGS field. Thirty-five targets from all the CEERS $z>8$ photometric samples in the literature have been observed with NIRSpec, leading to 25 robust redshift measurements. We measure a low fraction of lower-$z$ interlopers, with only two objects (8\%) identified as such. Such a high confirmation rate at $z>8$ reinforces the surprisingly high number densities and brightness of early galaxies compared to the theoretical predictions found by many photometric works in the literature.

Galaxy selection from deep \textit{JWST}/NIRCam imaging is essential to minimize selection bias at the highest redshifts.  The \textit{JWST}/NIRSpec prism is an optimal tool for measuring such high redshifts, but observations deeper than those of CEERS will be necessary to achieve a higher degree of spectroscopic completeness.

\begin{acknowledgments}

We thank the Space Telescope Science Institute (STScI) and GTO NIRSpec teams for their invaluable efforts to make the best out of such a wonderful spectrograph. In particular, we especially thank James Muzerolle, Alaina Henry, Patrick Ogle, Pierre Ferruit, Peter Jakobsen, Diane Karakla, Maria Peña-Guerrero, Amaya Moro-Martín and James Davies for their frequent assistance with the development of the CEERS NIRSpec observations. We also thank Rychard Bouwens, Callum Donnan and Ryan Endsley for providing updated samples of their early high-$z$ selections, and the anonymous referee for a very fast and positive review.

This work is based on observations with the NASA/ESA/CSA James Webb Space Telescope obtained from the Mikulski Archive for Space Telescopes at the STScI, which is operated by the Association of Universities for Research in Astronomy (AURA), Incorporated, under NASA contract NAS5-03127.
The specific observations analyzed can be accessed via \dataset[DOI: 10.17909/z7p0-8481]{https://doi.org/10.17909/z7p0-8481}.

We acknowledge support from NASA through STScI ERS award JWST-ERS-1345.

\end{acknowledgments}

\vspace{5mm}

\software{
Astropy \citep{Astropy2022};
Bagpipes \citep{Carnall2018};
Cigale \citep{Burgarella2005, Noll2009, Boquien2019};
Dense Basis \citep{Iyer2019}; 
emcee \citep{foreman-mackey13};
Galfit \citep{peng02, peng10};
LiMe \citep{Fernandez2023};
Mosviz \citep{JDADF2023}; 
Source Extractor \citep{Bertin1996}.
}

\newpage

\appendix

\vspace{-3.5mm}

\section{Complete census of $z>8$ spectroscopic follow-up in the CEERS field}
\label{sec:appendix_CEERS_spec_compilation}

Here we present a compilation of all photometric $z>8$ candidates in CEERS with spectroscopic follow-up in either the CEERS observations or the public DD \#2750 program. Table~\ref{tab:app_complete_zspec} presents all objects with robust spectroscopic measurements, while targets undetected or with uncertain spectroscopic measurements are shown in Table~\ref{tab:app_undetected_zspec}.


\begin{longtable*}{lccccccccc}
\caption{Complete sample of CEERS $z>8$ candidates with robust spectroscopic measurements.}
\label{tab:app_complete_zspec}
\\
\hline\hline
Source ID  &  MSA ID  & R.A.  &  Dec.  &  $M_{\mathrm{UV}}$  &  $m_{\mathrm{F277W}}$  &  Photo.  &  $z_\mathrm{phot}$ & $z_\mathrm{spec}$ &  Spec. \\
  &   &  (deg)  &  (deg)  &  (AB)  &  (AB)  &  Ref.  &    &    &  Ref. \\
(1)  &  (2)  &  (3)  &  (4)  &  (5)  &  (6)  &  (7)  &  (8)  &  (9)  &  (10) \\
\hline 
\endfirsthead
\multicolumn{10}{c}{\tablename\ \thetable\ -- \textit{Continued from previous page}}\\
\hline
Source ID  &  MSA ID  & R.A.  &  Dec.  &  $M_{\mathrm{UV}}$  &  $m_{\mathrm{F277W}}$  &  Photo.  &  $z_\mathrm{phot}$ & $z_\mathrm{spec}$ &  Spec. \\
  &   &  (deg)  &  (deg)  &  (AB)  &  (AB)  &  Ref.  &    &    &  Ref. \\
(1)  &  (2)  &  (3)  &  (4)  &  (5)  &  (6)  &  (7)  &  (8)  &  (9)  &  (10) \\
\hline 
\endhead
\hline \multicolumn{10}{r}{\textit{Continued on next page}} \\
\endfoot
\multicolumn{10}{l}{\parbox{\dimexpr\textwidth-\tabcolsep}{
(1) Source ID in the CEERS catalog. Galaxies out of the CEERS NIRCam imaging footprint are identified with their CANDELS EGS ID from \cite{Stefanon2017}.
(2) ID in the MSA observations. Those preceded by a letter ``D'' correspond to observations from the DD \#2750 program.
(3) Right ascension (J2000).
(4) Declination (J2000).
(5) Absolute UV magnitude measured at 1500~\AA{} rest-frame.
(6) Apparent magnitude in the NIRCam/F277W band.
(7) Reference for the photometric redshift: B23 \citep{Bouwens2023}; D23 \citep{Donnan2023}; E22 \citep{Endsley2022}; H23b \citep{Harikane2023a}; L23 \citep{Labbe2023}; W23 \citep{Whitler2023}. Objects labeled as F23 include both sources from \cite{Finkelstein2023} and new sources in the complete CEERS NIRCam coverage selected following the same criteria used in that work.
(8) Photometric redshift.
(9) Spectroscopic redshift remeasured in this work for emission lines redshifts following the methodology described in \S\ref{sec:redshift}.
(10) Works spectroscopically reporting these sources: AH23 \citep{ArrabalHaro2023}; Fu23 \citep{Fujimoto2023}; He23 \citep{Heintz2023}; I23 \citep{Isobe2023}; J22 \citep{Jung2022}; J23 \citep{Jung2023}; K23 \citep{Kocevski2023}; La23 \citep{Larson2023}; N23 \citep{Nakajima2023}; S23 \citep{Sanders2023}; T23 \citep{Tang2023}; Z15 \citep{Zitrin2015}.
}}\\
\hline
\endlastfoot
\hline
     CEERS\_16943  &     D1  &  214.943152  &   52.942442  &  $-20.1^{+0.1}_{-0.1}$  &  27.9  &  B23  &  $11.60^{+0.40}_{-0.50}$  &  $11.416^{+0.005}_{-0.005}$  &  AH23  \\
                   &         &              &              &                       &        &  D23  &  $12.29^{+0.91}_{-0.32}$  &                          &    \\
                   &         &              &              &                       &        &  F23  &  $11.08^{+0.39}_{-0.36}$  &                          &    \\
                   &         &              &              &                       &        &  H23b  &  $11.63^{+0.51}_{-0.53}$  &                          &    \\
\hline
     CEERS\_11384  &    D10  &  214.906640  &   52.945504  &  $-20.3^{+0.1}_{-0.2}$  &  27.3  &  D23  &  $11.27^{+0.58}_{-0.27}$  &  $11.043^{+0.003}_{-0.003}$  &  AH23,H23a  \\
                   &         &              &              &                       &        &  F23  &  $11.02^{+0.39}_{-0.27}$  &                          &    \\
\hline
     CEERS\_19996  &    D64  &  214.922787  &   52.911529  &  $-19.3^{+0.2}_{-0.2}$  &  28.3  &  D23  &  $11.27^{+0.39}_{-0.70}$  &  $10.10^{+0.13}_{-0.26}$  &  AH23  \\
                   &         &              &              &                       &        &  F23  &  $11.32^{+0.30}_{-0.90}$  &                          &    \\
\hline
     CEERS\_35590  &  80041  &  214.732525  &   52.758090  &  $-20.1^{+0.1}_{-0.1}$  &  27.7  &  F23  &  $10.15^{+0.36}_{-0.42}$  &  $10.01^{+0.14}_{-0.19}$  &  This work  \\
\hline
     CEERS\_99715  &  80026  &  214.811852  &   52.737110  &  $-20.5^{+0.1}_{-0.1}$  &  27.1  &  F23  &  $ 9.76^{+0.60}_{-0.09}$  &  $ 9.77^{+0.37}_{-0.29}$  &  This work  \\
\hline
     CEERS\_61419  &     24  &  214.897232  &   52.843854  &  $-19.3^{+0.2}_{-0.1}$  &  28.1  &  B23  &  $10.20^{+0.60}_{-0.70}$  &  $ 8.998^{+0.001}_{-0.001}$  &  Fu23,T23  \\
                   &         &              &              &                       &        &  D23  &  $10.56^{+0.25}_{-0.52}$  &                          &    \\
                   &         &              &              &                       &        &  F23  &  $ 8.95^{+1.65}_{-0.06}$  &                          &    \\
                   &         &              &              &                       &        &  W23  &  $ 8.95^{+0.07}_{-0.09}$  &                          &    \\
\hline
     CEERS\_61381  &     23  &  214.901252  &   52.846997  &  $-18.9^{+0.2}_{-0.1}$  &  28.5  &  D23  &  $10.45^{+0.52}_{-0.96}$  &  $ 8.881^{+0.001}_{-0.001}$  &  Fu23,T23  \\
                   &         &              &              &                       &        &  F23  &  $11.29^{+0.21}_{-1.56}$  &                          &    \\
\hline
      CEERS\_7078  &      7  &  215.011708  &   52.988303  &  $-20.6^{+0.1}_{-0.1}$  &  27.1  &  F23  &  $ 8.98^{+0.06}_{-0.06}$  &  $ 8.876^{+0.002}_{-0.002}$  &  Fu23,N23  \\
                   &         &              &              &                       &        &  W23  &  $ 9.00^{+0.05}_{-0.06}$  &                          &    \\
\hline
      CEERS\_4702  &      2  &  214.994404  &   52.989378  &  $-20.2^{+0.1}_{-0.1}$  &  27.5  &  B23  &  $ 9.20^{+0.10}_{-0.20}$  &  $ 8.809^{+0.003}_{-0.003}$  &  Fu23  \\
                   &         &              &              &                       &        &  F23  &  $ 8.98^{+0.12}_{-0.12}$  &                          &    \\
                   &         &              &              &                       &        &  W23  &  $ 8.92^{+0.09}_{-0.09}$  &                          &    \\
\hline
     CEERS\_43833  &    D28  &  214.938642  &   52.911749  &  $-20.7^{+0.1}_{-0.1}$  &  26.8  &  F23  &  $ 9.01^{+0.09}_{-0.09}$  &  $ 8.763^{+0.001}_{-0.001}$  &  AH23  \\
\hline
     CEERS\_43725  &   1025  &  214.967532  &   52.932953  &  $-21.3^{+0.1}_{-0.1}$  &  26.3  &  F23  &  $ 8.68^{+0.06}_{-0.09}$  &  $ 8.715^{+0.001}_{-0.001}$  &  He23,T23  \\
\hline
     CEERS\_81061  &   1019  &  215.035392  &   52.890667  &  $-22.2^{+0.1}_{-0.1}$  &  25.0  &  F23  &  $ 8.68^{+0.06}_{-0.03}$  &  $ 8.679^{+0.001}_{-0.001}$  &  He23,I23,La23  \\
        &      &     &      &     &     &     &     &     &  S23,T23,Z15  \\
\hline
     CEERS\_90671  &  80083  &  214.961276  &   52.842364  &  $-18.7^{+0.1}_{-0.1}$  &  28.1  &  F23  &  $ 8.68^{+0.21}_{-0.27}$  &  $ 8.638^{+0.001}_{-0.001}$  &  This work  \\
\hline
EGS\_11855  &   1029  &  215.218762  &   53.069862  &  ---  &   ---  &  F23  &  $ 8.95^{+0.15}_{-0.43}$  &  $ 8.610^{+0.001}_{-0.001}$  &  He23,T23  \\
\hline
EGS\_34697  &   1149  &  215.089714  &   52.966183  &  ---  &   ---  &  F23  &  $ 6.45^{+0.68}_{-4.82}$  &  $ 8.175^{+0.001}_{-0.001}$  &  He23,S23,T23  \\
\hline
      CEERS\_4774  &      3  &  215.005185  &   52.996577  &  $-19.6^{+0.3}_{-0.3}$  &  27.0  &  F23  &  $ 8.92^{+1.35}_{-0.66}$  &  $ 8.005^{+0.001}_{-0.001}$  &  Fu23,N23,T23  \\
\hline
      CEERS\_4777  &      4  &  215.005365  &   52.996697  &  $-18.7^{+0.5}_{-0.2}$  &  28.0  &  B23  &  $ 9.90^{+0.80}_{-0.80}$  &  $ 7.993^{+0.001}_{-0.001}$  &  Fu23  \\
                   &         &              &              &                       &        &  F23  &  $10.12^{+0.93}_{-0.69}$  &                          &    \\
                   &         &              &              &                       &        &  L23  &  $ 8.62^{+0.34}_{-0.57}$  &                          &    \\
\hline
     CEERS\_19185  &   D355  &  214.944766  &   52.931450  &  $-19.0^{+0.1}_{-0.1}$  &  28.7  &  F23  &  $ 8.20^{+0.33}_{-0.24}$  &  $ 7.925^{+0.001}_{-0.001}$  &  AH23  \\
\hline
     CEERS\_59920  &   1027  &  214.882994  &   52.840416  &  $-20.8^{+0.1}_{-0.1}$  &  26.5  &  E22  &  $ 7.82^{+0.04}_{-0.03}$  &  $ 7.820^{+0.001}_{-0.001}$  &  He23,S23,T23  \\
                   &         &              &              &                       &        &  F23  &  $ 8.17^{+0.06}_{-0.12}$  &                          &    \\
\hline
EGS\_8901  &   1023  &  215.188413  &   53.033647  &  ---  &   ---  &  F23  &  $ 8.85^{+0.18}_{-1.15}$  &  $ 7.776^{+0.001}_{-0.001}$  &  He23,T23  \\
\hline
     CEERS\_23084  &     20  &  214.830685  &   52.887771  &  $-17.6^{+0.1}_{-0.6}$  &  28.2  &  F23  &  $ 8.77^{+0.45}_{-0.69}$  &  $ 7.769^{+0.003}_{-0.003}$  &  Fu23  \\
                   &         &              &              &                       &        &  L23  &  $ 9.08^{+0.31}_{-0.38}$  &                          &    \\
\hline
EGS\_33634  &    686  &  215.150862  &   52.989562  &  ---  &   ---  &  F23  &  $ 8.00^{+0.52}_{-1.03}$  &  $ 7.752^{+0.001}_{-0.001}$  &  J22,J23,T23  \\
\hline
     CEERS\_96937  &  80025  &  214.806065  &   52.750867  &  $-20.0^{+0.2}_{-0.1}$  &  27.4  &  F23  &  $ 8.47^{+0.15}_{-0.24}$  &  $ 7.651^{+0.002}_{-0.001}$  &  This work  \\
\hline
EGS\_36986  &    689  &  214.999053  &   52.941977  &  ---  &   ---  &  F23  &  $ 8.61^{+0.11}_{-1.45}$  &  $ 7.546^{+0.001}_{-0.001}$  &  J22,J23,T23  \\
\hline
     CEERS\_24015  &    746  &  214.809155  &   52.868481  &  $-14.1^{+1.1}_{-0.7}$  &  28.1  &  L23  &  $ 8.14^{+0.45}_{-1.71}$  &  $ 5.623^{+0.001}_{-0.001}$  &  K23  \\
\hline
     CEERS\_13256  &     D0  &  214.914550  &   52.943023  &  $-16.2^{+0.5}_{-0.1}$  &  26.5  &  B23  &  $16.30^{+0.30}_{-0.40}$  &  $ 4.912^{+0.001}_{-0.001}$  &  AH23  \\
                   &         &              &              &                       &        &  D23  &  $16.39^{+0.32}_{-0.22}$  &                          &    \\
                   &         &              &              &                       &        &  F23  &  $16.45^{+0.18}_{-0.45}$  &                          &    \\
                   &         &              &              &                       &        &  H23b  &  $16.25^{+0.24}_{-0.46}$  &                          &    \\

\hline \hline

\end{longtable*}



\begin{longtable*}{lcccccc}
\caption{Complete sample of CEERS $z>8$ candidates in NIRSpec MSA observations without robust spectroscopic redshift. See description of columns in Table~\ref{tab:app_complete_zspec}.}
\label{tab:app_undetected_zspec}
\\
\hline\hline
Source ID  &  MSA ID  & R.A.  &  Dec.   &  $m_{\mathrm{F277W}}$  &  Photo.  &  $z_\mathrm{phot}$ \\
  &   &  (deg)  &  (deg)  &  (AB)  &  Ref.  &     \\
(1)  &  (2)  &  (3)  &  (4)  &  (5)  &  (6)  &  (7) \\
\hline 
\endfirsthead
\multicolumn{7}{c}{\tablename\ \thetable\ -- \textit{Continued from previous page}}\\
\hline
Source ID  &  MSA ID  & R.A.  &  Dec.   &  $m_{\mathrm{F277W}}$  &  Photo.  &  $z_\mathrm{phot}$ \\
  &   &  (deg)  &  (deg)  &  (AB)  &  Ref.  &     \\
(1)  &  (2)  &  (3)  &  (4)  &  (5)  &  (6)  &  (7) \\
\hline 
\endhead
\hline \multicolumn{7}{r}{\textit{Continued on next page}} \\
\endfoot
\hline
\endlastfoot
\hline 
CEERS\_2067   &      0  &  215.010022  &   53.013641  &  27.8  &  F23  &  $13.69^{+0.66}_{-0.99}$  \\
\hline
CEERS\_49703  &    733  &  214.910343  &   52.855032  &  28.4  &  D23  &  $11.90^{+0.70}_{-1.60}$  \\
\hline
CEERS\_87379  &  80073  &  214.932064  &   52.841873  &  27.3  &  F23  &  $11.08^{+0.24}_{-0.48}$  \\
\hline
CEERS\_10332  &      9  &  215.043999  &   52.994302  &  28.4  &  D23  &  $10.80^{+0.51}_{-0.40}$  \\
              &         &              &              &        &  F23  &  $10.57^{+0.18}_{-1.05}$  \\
\hline
CEERS\_57400  &     22  &  214.869661  &   52.843646  &  28.7  &  F23  &  $10.60^{+0.60}_{-0.66}$  \\
\hline
CEERS\_13452  &    D69  &  214.861602  &   52.904604  &  28.2  &  D23  &  $10.56^{+1.10}_{-0.30}$  \\
              &         &              &              &        &  F23  &  $ 9.55^{+0.78}_{-0.09}$  \\
\hline
CEERS\_36952  &  80099  &  214.771106  &   52.780817  &  29.1  &  F23  &  $10.48^{+6.00}_{-0.81}$  \\
\hline
CEERS\_42447  &  80042  &  214.795552  &   52.767286  &  28.3  &  F23  &  $10.30^{+0.09}_{-1.11}$  \\
\hline
CEERS\_4821   &    853  &  215.012542  &   53.001372  &  27.5  &  W23  &  $ 9.16^{+0.06}_{-0.06}$  \\
\hline
CEERS\_56878  &     21  &  214.888127  &   52.858987  &  27.7  &  F23  &  $ 9.01^{+0.30}_{-0.30}$  \\

\hline \hline

\end{longtable*}

\section{Measuring and rescaling noise in the NIRSpec spectra}
\label{sec:appendix_noise}

The \textit{JWST} pipeline resamples the two-dimensional NIRSpec MSA spectra in order to align and combine the individual nodded exposures and to rectify the two-dimensional spectra so that the dispersion and cross-dispersion axes align with \textit{x} and \textit{y} pixel coordinates in the {\tt s2d} pipeline data product. The pipeline then extracts a one-dimensional spectrum ({\tt x1d}) from the {\tt s2d} product, including \texttt{FLUX} and \texttt{FLUX\_ERROR} values.  The flux errors are calculated by the pipeline using an instrumental noise model, and the resampling introduces correlation between pixels that smooth the data, reducing the RMS of the measured pixel values. 

We test the flux errors in the {\tt x1d} spectra for our faint galaxy targets CEERS\_99715 and CEERS\_35590 by dividing the fluxes by the flux errors to calculate a signal-to-noise ratio ($S/N$) spectrum. These galaxies are very faint, without strong emission lines or other features so that \texttt{FLUX\_ERROR} should be dominated by background and instrument noise. After masking pixel artifacts and missing data (\S\ref{sec:NIRSpec_reduction}) we smooth the $S/N$ spectrum by a 30 pixel boxcar filter and subtract to produce a noise residual spectrum with zero mean and approximately constant dispersion over the full wavelength range. If noise in the data has a Gaussian distribution and is correctly characterized by the pipeline's \texttt{FLUX\_ERROR} values, the values of this noise residual spectrum should be Gaussian with $\sigma = 1$. The noise residuals for both objects show nearly Gaussian distributions with $\sigma_{\mathrm{S/N}} = 1.30$ and 1.24 for objects CEERS\_99715 and CEERS\_35590, respectively.

We also measure the autocorrelation function $\xi(\delta x)$ of the $S/N$ spectrum as a function of pixel lag $\delta x$.  For uncorrelated random noise, $\xi$ would be a delta function at $\delta x = 0$. Instead, we observe positive correlation on a scale of $|\delta x| = 1$ to 2 pixels.  The 30 pixel boxcar subtraction also introduces anticorrelation out to scales of $|\delta x| = 3$ to 15 pixels.  We calculate the excess correlation over this negative ``background'' measured at $|\delta x|= 3$~pixels:
\[X = \sum_{\delta x=-2}^{+2} (\xi(\delta x) - \xi(|\delta x|=3))\]
For noise with intrinsic RMS = $\sigma_0$, interpolation reduces the apparent RMS = $\sigma_{\mathrm{obs}}$ by a factor $F$:
\[\sigma_{\mathrm{obs}} = \sigma_0 / F\]
\[F = \sqrt{X/\xi(0)}\]
We measure $F = 1.42$ and 1.33 for objects CEERS\_99715 and CEERS\_35590.  Multiplying $F$ by the RMS rescaling factors $\sigma_{\mathrm{S/N}}$ calculated above, we derive total error correction factors of 1.85 and 1.65 for these two objects, respectively, and multiply the pipeline \texttt{FLUX\_ERROR} values by these factors.  We expect the noise scaling and correlation to be similar for the prism spectra of all faint galaxy targets, and therefore adopt a noise rescaling factor of 1.75 for the other objects analyzed in this paper.  

\section{Caveat on the robustness of the photometric selection: Possible interlopers in relaxed photometric samples}
\label{sec:appendix_possible_interlopers}

We note here that the primary photometric sample of seven $z > 8$ galaxies was selected via a robust set of photometric detection and photometric-redshift-based selection criteria.  As fully described in \citet{Finkelstein2023}, our photometric redshift selection makes use of several integrals of the photometric redshift $P(z)$, rather than relying solely on the best-fitting redshift.

\begin{figure*}
\centering
\includegraphics[width=\textwidth]{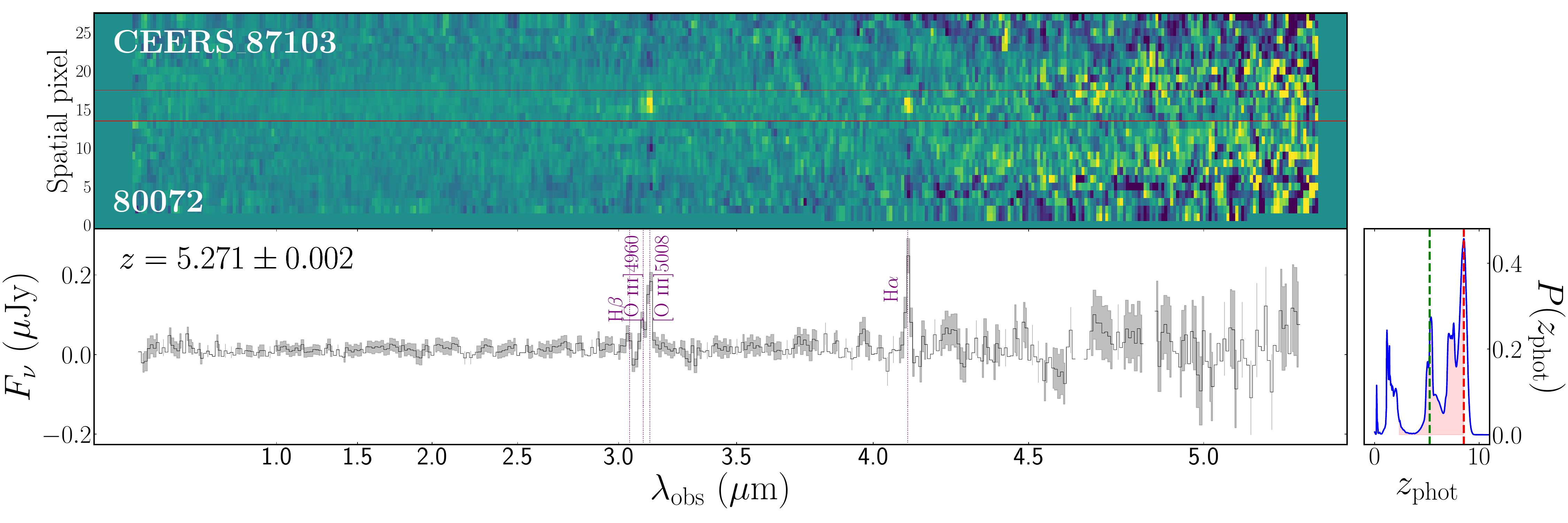}
\caption{1D (\textit{bottom left}) and 2D (\textit{top left}) spectrum of CEERS\_87103. The two red horizontal lines in the 2D spectrum show the spatial window employed to extract the 1D spectrum and the flux uncertainties of the 1D spectrum are represented by the shaded grey regions. The broad photometric redshift probability distribution function $P(z)$ obtained for this source (\textit{right}) peaks at $z=8.53$ (dashed red line), which could result in the inclusion of an object like this in high-$z$ samples under relaxed selection criteria. The shaded red region behind the $P(z)$ shows the 68\% confidence interval of the photometric redshift estimation, with the dashed green line indicating the spectroscopic value.}
\label{fig:possible_interloper}
\end{figure*}

Fig.~\ref{fig:possible_interloper} shows the 1D and 2D spectrum of one of these objects, CEERS\_87103 (MSA ID 80072), observed in CEERS epoch 3.   The spectra reveals an unambgious redshift of $z =$ 5.27 based on the H$\alpha$ and \oiii\ emission lines. The right-hand panel shows its $P(z)$ peaking at $z \sim$ 8, which led to our placing of a slit on it.  However, this distribution is also quite broad, with several peaks at lower redshift.  This galaxy was labeled as a less-secure high-$z$ candidate for two specific reasons.  First, the integral of the $P(z)$ at $z >$ 7 was 0.55, less than the threshold of 0.7 adopted by \citet{Finkelstein2023}.  Second, the difference in the goodness-of-fit $\chi^2$ value between the primary $z \sim$ 8 peak and the highest lower-redshift, $z \sim$ 5, peak was 1.5, much less than the threshold of four adopted by \citet{Finkelstein2023}.  This highlights the importance of employing robust selection criteria in defining high-$z$ samples, as more relaxed high-$z$ selections based solely on the best-fitting redshift can be prone to include lower $z$ interlopers in the selected samples.

\bibliography{biblio}{}
\bibliographystyle{aasjournal}



\end{document}